\documentclass[twocolumn,showpacs,preprintnumbers,
amsmath,amssymb,floatfix,nofootinbib]{revtex4-1}

\usepackage{dcolumn}
\usepackage{bm}
\usepackage{amsthm}
\usepackage{amsfonts}

\usepackage{anysize}

\usepackage{amsmath}
\usepackage{mathtools}
\usepackage[english]{babel}
\usepackage{graphicx}

\usepackage{caption}
\usepackage{subcaption}

\usepackage{amssymb}

\newcommand{\be}{\begin{equation}}
\newcommand{\ee}{\end{equation}}
\newcommand{\ba}{\begin{eqnarray}}
\newcommand{\ea}{\end{eqnarray}}
\newcommand{\non}{\nonumber}

\newcommand{\ce}{{\cal{E}}}
\newcommand{\cl}{{\cal{L}}}
\newcommand{\ck}{{\cal{K}}}

\newcommand{\nin}{\noindent}

\usepackage{color}

\begin{document}

\title{Escape of Charged Particles Moving around a Weakly Magnetized Kerr Black Hole}
\author{A. M. Al Zahrani}
\email{ama3@ualberta.ca}
\affiliation{Theoretical Physics Institute, University of Alberta, Edmonton, Alberta T6G 2E1, Canada}

\begin{abstract}
  We study the dynamics and escape of charged particles initially orbiting a weakly magnetized Kerr black hole after they get kicked in the direction normal to the orbit. 
  The case of neutral particles is analysed first and the escape conditions are given analytically. A general analysis of charged particles innermost stable circular orbits (ISCO)s is performed numerically. We then study the charged particles three-dimensional motion and give an effective condition for their escape. We also discuss how the black hole's rotation affects the escape of charged particles and the chaoticness in their dynamics.
\end{abstract}


\maketitle

\section{Introduction}

The dynamics of bipolar jets observed near astrophysical black holes and active galactic nuclei in particular, remains a mystery. There have been several jet launching and collimation mechanisms proposed, which involve magnetic fields as an essential ingredient. Nowadays, the problem is most commonly approached via the advanced computer simulations of the gravitohydromagnetics of plasma accreting into rotating black holes. (See, e.g., Ref.~\cite{RoVi} and the references therein.) It is unknown whether the jets are powered by the accretion disk or the rotational energy of the black hole. Recent observations concluded that the power of the jets is proportional to the black hole's spin, in agreement with the mechanism proposed by Blandford and Znajek \cite{Nar0,Nar}. This conclusion is congruent with that of computer simulations of the gravitohydromagnetics (see, e.g., Refs.~\cite{KSKM,SDP,TMN}). It should be noted, however, that a pervious observation found no evidence for black hole rotation powering the jets in x-ray binaries \cite{Fend}.

Magnetic fields can be present in the vicinity of a black hole, mainly due to the accreting plasma around it as discussed in Refs. \cite{RoVi,Pun}. Moreover, astrophysical black holes are speculated to be rapidly rotating. Even slowly rotating black holes can be spun up by matter accretion \cite{Bar,LB}. The spin angular momentum of a black hole of mass $M$ is thought to be limited by $J=0.998M^2$ due to the counteracting torque resulting from the absorption of the radiation from the accretion disk \cite{Th}. Recent observations found that astrophysical black holes are indeed rapidly rotating~\cite{Bren,McC0,Rey,McC}.

In this paper we consider a simplified and yet interesting model that can shed light on the high energy emissions associated with astrophysical black holes. The system we study consists of a charged particle in a circular orbit around a rotating black hole immersed in a uniform weak axisymmetric magnetic field. The field is weak in the sense that its back-reaction on the spacetime is negligible. The field is either aligned or oppositely aligned with the black hole's spin.

We then give the particle a kick off the orbit and observe how its dynamics evolves and whether it escapes or ends up captured by the black hole. In real situations the kick could be given for example by another particle or photon. The problem in the background of a Schwarzschild black hole was studied in Ref.~\cite{Z2}.

The inclusion of the magnetic field breaks down the constant of motion associated with the Kerr spacetime's hidden symmetry; the Carter constant. Consequently, the equations of motion are rendered non-integrable in general. They remain integrable in the equatorial submanifold, however. The main effect of the magnetic field on the charged particles' circular orbits is bringing their ISCOs closer to the black hole. Additionally, negatively 'superbound' stable circular orbits can exist if the magnetic force is large enough (see below).

Numerical integration is required for studying the dynamics outside the equatorial submanifold. Depending on the initial conditions and the parameters of the system, the motion can be chaotic. The chaotic motion of charged particles near a Kerr black hole immersed in a weak magnetic field was studied in Refs.~\cite{kkss,kkss2} for a uniform axisymmetric field and in Refs.~\cite{ni,tk} for a dipole field. Similar studies in the background of a Schwarzschild black hole were conducted. In fact, there are several cases in general relativity where chaotic particle dynamics was encountered even in the absence of magnetic fields. (See the references in Ref.~\cite{Z2})

In this paper we study charged particles escape from a weakly magnetized rotating black hole. The simpler case of neutral particles is tackled first. The effect of the black hole's rotation on charged particles escape and chaoticness in their dynamics is investigated as well. The paper is organised as follows: In Sec.~\ref{s2} we analyse the case of neutral particles. We review particle dynamics and circular orbits in Kerr geometry and then give the escape conditions analytically. In Sec.~\ref{s3} we treat the charged particles case. We introduce the magnetization of rotating black holes, describe circular orbits and ISCOs, and then analyse charge particles dynamics and give the conditions for  their escape. The relationship between chaoticness and rotation is investigated afterward. We give general discussion and conclusion in Sec.~\ref{sum}. We use the sign conventions
adopted in Ref.~\cite{MTW} and geometrical units where $c=G=1$.

\section{Escape Velocity of a Neutral Particle} \label{s2}

\subsection{Circular Orbits}

The spacetime geometry around a rotating black hole is described by the Kerr metric. For a black hole of mass $M$ and spin angular momentum $J=aM$ the Kerr metric in Boyer-Linquist coordinates reads~\cite{Zee}
\begin{eqnarray}
ds^2=&-&\Sigma\frac{\Delta}{A}dt^2+\frac{\Sigma}{\Delta}dr^2+\Sigma d\theta^2\nonumber \\
&+&\frac{A}{\Sigma}\left(d\phi-\frac{2aMr}{A}dt\right)^2\sin^2\theta,
\end{eqnarray}
where
\begin{eqnarray}
\Sigma &=&r^2+a^2\cos^2\theta, \:\:\: \Delta=r^2+a^2-2Mr, \non \\
&&\:\:\:A=(r^2+a^2)^2-a^2\Delta \sin^2\theta,
\end{eqnarray}

\nin and $a$, with $-M\leq a\leq M$, is the rotation parameter.

The Kerr spacetime admits two commuting Killing vectors
\begin{equation}
\xi^{\mu}_{(t)}=\delta^{\mu}_t, \:\:\: \xi^{\mu}_{(\phi)}=\delta^{\mu}_{\phi},
\end{equation}

\noindent and a Killing tensor
\begin{equation}
K^{\mu\nu}=\Delta k^{(\mu}l^{\nu)}+r^2g^{\mu\nu},
\end{equation}

\noindent where
\begin{eqnarray}
l^{\mu}&=&\frac{1}{\Delta}\left[(r^2+a^2)\delta_t^{\mu}+\Delta \delta_r^{\mu}+a \delta_{\phi}^{\mu}\right], \\
k^{\mu}&=&\frac{1}{\Delta}\left[(r^2+a^2)\delta_t^{\mu}-\Delta \delta_r^{\mu}+a \delta_{\phi}^{\mu}\right].
\end{eqnarray}

\noindent Consider a particle in the Kerr spacetime moving with four-velocity $u^{\mu}$. The three Killing symmetries are associated with three constants of the particle's motion
\begin{eqnarray}
-\ce &=& p_{\mu}\xi^{\mu}_{(t)}/m,\\
\cl &=& p_{\mu}\xi^{\mu}_{(\phi)}/m,\\
\ck &=& u_{\mu}u_{\nu}K^{\mu\nu}-(\cl-a\ce)^2\label{car},
\end{eqnarray}

\noindent where $p^{\mu}=mu^{\mu}$ is the particle's four-momentum. $\ce$ and $\cl$ are the specific energy and azimuthal angular momentum, respectively, and $\ck$ is the Carter constant\footnote{The second term on the right hand side of Eq.~(\ref{car}) does not appear in the standard definition of the Carter constant. We chose our definition for convenience.}. Using these three constants of motion along with the normalization $u_{\mu}u^{\mu}=-1$ we reduce the equations of motion to quadratures:
\begin{eqnarray}
&&\dot{t}=\ce+\frac{2Mr[(r^2+a^2)\ce-a\cl]}{\Delta\Sigma},\\
&&\dot{\phi}=\frac{\cl}{\Sigma\sin^2\theta}+\frac{a(2Mr\ce-a\cl)}{\Delta\Sigma},\\
&&\Sigma^2\dot{r}^2=[(r^2+a^2)\ce-a\cl]^2 \nonumber\\
&&\hspace{1.2cm}-\Delta[r^2+\ck+(\cl-a\ce)^2],\label{rdot}\\
&&\Sigma^2\dot{\theta}^2=\ck+(\cl-a\ce)^2-a^2\cos^2{\theta} \nonumber\\
&&\hspace{1.2cm}-\left(a\ce\sin{\theta}-\frac{\cl}{\sin{\theta}}\right)^2\label{thethadot},
\end{eqnarray}

\noindent where the overdot denotes the derivative with respect to the proper time. The dynamics is invariant under reflection with respect to the equatorial plane
\begin{equation}\label{sym1}
\theta\rightarrow\pi-\theta, \hspace{1cm} \dot{\theta}\rightarrow-\dot{\theta}.
\end{equation}
It is also invariant under the transformations
\begin{equation}\label{sym2}
\phi\rightarrow-\phi, \hspace{3mm} \dot{\phi}\rightarrow-\dot{\phi}, \hspace{3mm} \cl\rightarrow-\cl, \hspace{3mm} a\rightarrow -a.
\end{equation}
There are two dynamically distinct modes of motion, depending on whether the black hole's spin and particle's azimuthal angular momentum are aligned ($a\cl>0$) or oppositely aligned ($a\cl<0$). Without loss of generality, $\cl$ will be kept positive while $a$ can take both signs. We refer to orbits with $a>0$ as {\it prograde} and orbits with $a<0$ as {\it retrograde}.

Let us define $R(r)$ to be the right hand side of Eq.~(\ref{rdot}):  
\begin{eqnarray}
R(r):=&[(&r^2+a^2)\ce-a\cl]^2 \nonumber\\
&-&\Delta[r^2+\ck+(\cl-a\ce)^2].\label{rpot}
\end{eqnarray}
$R(r)$ is positive semidefinite; it vanishes at the radial turning points only. Equatorial circular orbits exist where $R(r)$ and its first derivative $R'(r)$ vanish when $\theta=\frac{\pi}{2}, \ck=0$. We used the notation $(\:\:\:)'=\partial_r(\:\:\:)$. These two conditions yield
\begin{eqnarray}
&&[(r^2+a^2)\ce-a\cl]^2-\Delta[r^2+(\cl-a\ce)^2]=0,\:\:\:\:\:\:\: \\
&&2r\ce[(r^2+a^2)\ce-a\cl]-2r\Delta \nonumber\\
&&\hspace{18mm}-2(r-M)[r^2+(\cl-a\ce)^2]=0.\:\:
\end{eqnarray}

\nin We will use $r_o$, $\ce_o$ and $\cl_o$ to denote quantities corresponding to circular orbits from here on. Solving these equations for $\ce_o$ and $\cl_o$ one obtains
\begin{eqnarray}
\ce_o=\frac{aM^{1/2}+r_o^{1/2}(r_o-2M)}{\sqrt{2aM^{1/2}r_o^{3/2}+r_o^2(r_o-3M)}},\label{ec}\\
\cl_o=\frac{M^{1/2}(a^2+r_o^2)-2aMr_o^{1/2}}{\sqrt{2aM^{1/2}r_o^{3/2}+r_o^2(r_o-3M)}}.\label{lc}
\end{eqnarray}

\nin The radius of the last circular orbit $r_{lc}$, is given by
\begin{equation}
r_{lc}=\frac{[M+M^{1/3}(\sqrt{a^2-M^2}-a)^{2/3}]^2}{M^{1/3}(\sqrt{a^2-M^2}-a)^{2/3}}.
\end{equation}

\nin Equation~(\ref{ec}) reveals that $\ce_o$ is positive for all circular orbits. A circular orbit is the ISCO when $R''(r_o)$ vanishes, or
\begin{eqnarray}
(6r_o^2+a^2)(\ce_o^2-1)+6Mr_o-\cl_o^2=0.
\end{eqnarray}

\nin Plugging the $\ce_o$ and $\cl_o$ expressions above in this condition yields
\begin{equation}
r_{ms}(r_{ms}-6M)+8a\sqrt{Mr_{ms}}-3a^2=0.
\end{equation}

\nin We used $r_{ms}$ (for {\it marginally stable}) to denote the ISCO's radius. The $\ce_o$ and $\cl_o$ expressions reduce for the ISCO to
\begin{eqnarray}
\cl_{ms}^2=\frac{2}{3}\frac{M}{r_{ms}}(3r_{ms}^2-a^2),\:\:\: \ce_{ms}^2=1-\frac{2}{3}\frac{M}{r_{ms}}.\:\:\:\:\:\:\:\:
\end{eqnarray}

\begin{figure}[h!]
  \centering
  \includegraphics[width=0.48\textwidth]{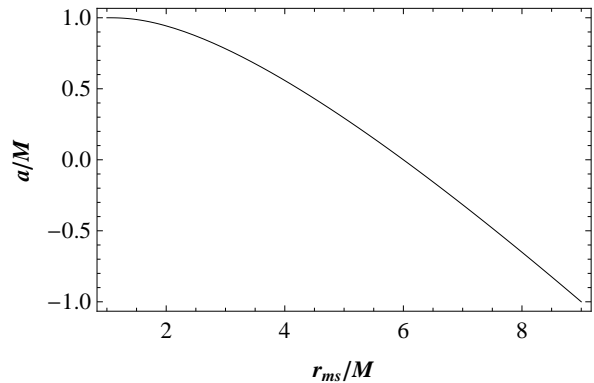}
  \caption{The dependence of the radius of the last stable circular orbit $r_{ms}$ on the black hole's rotation parameter $a$.}
  \label{fig:arms}
\end{figure}

\nin Figure~\ref{fig:arms} shows how $r_{ms}$ changes with $a$. The ISCO radius lies in the interval $[M,9M]$.

\subsection{Conditions for escape from a circular orbit}

\subsubsection{Three-Dimensional Motion}

A particle at a stable circular orbit of radius $r_o$ has the four-velocity
\begin{equation}
\tilde{u}_{\mu}=(-\ce_o,0,0,\cl_o).
\end{equation}
To reduce the complexity of the problem we will consider a kick that gives the particle polar velocity $v_k=-r_o\dot{\theta}_k$ without changing $\cl_o$. The kick therefore changes the particle's four-velocity to
\begin{equation}
u_{\mu}=(-\ce,0,r_o^3\dot{\theta}_k,\cl_o).
\end{equation}
The space of initial conditions of the problem is therefore two-dimensional: $\{r_o,\dot{\theta}_k\}$. The symmetry transformations~(\ref{sym1}) make it enough to take $v_k$ to be positive or negative without loss of generality. We can express the dependence of $\ce$ and $\ck$ on $\dot{\theta}_k$ using Eqs.~(\ref{rdot}) and~(\ref{thethadot}). The expressions are
{\small
\begin{eqnarray}
&{\cal E}&=\frac{1}{r_o^3+a^2(r_o+2M)}\Big[2a\cl_o M+\Delta_o^{1/2} \nonumber\\
& &\sqrt{a^2(r+2M)(r_o^3\dot{\theta}_k^2+r_o)+r_o^2(r^4_o\dot{\theta}_k^2+r^2_o+\cl_o^2)}\Big],\:\:\:\:\:\:\:\:\:\label{ene0}\\
&{\cal K}&=r_o^4\dot{\theta}_k^2, \label{cc}
\end{eqnarray}}where $\Delta_o=\Delta|_{r=r_o}$. The root for $\ce$ corresponding to future-directed four-velocity was selected.

To study the particle's behavior after the kick, it is more appropriate to recast Eq.~(\ref{rdot}) as
\begin{equation}\label{epe}
\Sigma^2\dot{r}^2=r[r^3+a^2(r+2M)](\ce-V_+)(\ce-V_-),
\end{equation}

\noindent where
\begin{eqnarray}
\small &V&_{\pm}(r)=\frac{1}{r^3+a^2(r+2M)}\Big[2a\cl M\pm\Delta^{1/2} \nonumber\\
& &\sqrt{a^2(r+2M)(\ck/r+r)+r^2(\ck+r^2+\cl^2)}\Big].\:\:\:\:\:\:\:\:\label{ene}
\end{eqnarray}
Again $V_+(r)$ will be considered for future-directed four-velocity vector. In order to determine the escape conditions we need to inspect $V_+(r)$ to figure out how the particle moves after getting kicked.

\subsubsection{Escape Conditions}

Far away from the black hole, $V_+(r)$ becomes unity. Trivially, the particle must be energetically unbound (${\cal E}\geq 1$) to be able to escape. The value of $\dot{\theta}_k$ at which the particle becomes energetically unbound is designated as $\dot{\theta}_{\ce=1}$. We use Eqs.~(\ref{ene0}) and~(\ref{cc}) to express it as
\begin{equation}
|\dot{\theta}_{\ce=1}|=\left[\frac{2M[(\cl_o-a)^2+r_o^2]-\cl_o^2r_o}{\Delta r_o^3}\right]^{1/2}.
\end{equation}
We will assume that the trivial condition $|\dot{\theta}_k|\ge |\dot{\theta}_{\ce=1}|$ is always satisfied. When $|\dot{\theta}_k|\ll|\dot{\theta}_{\ce=1}|$, the particle oscillates slightly around the initial orbit.

\begin{figure*}[ht]
\begin{center}
\ba
&&\hspace{0cm}\includegraphics[width=0.48\textwidth]{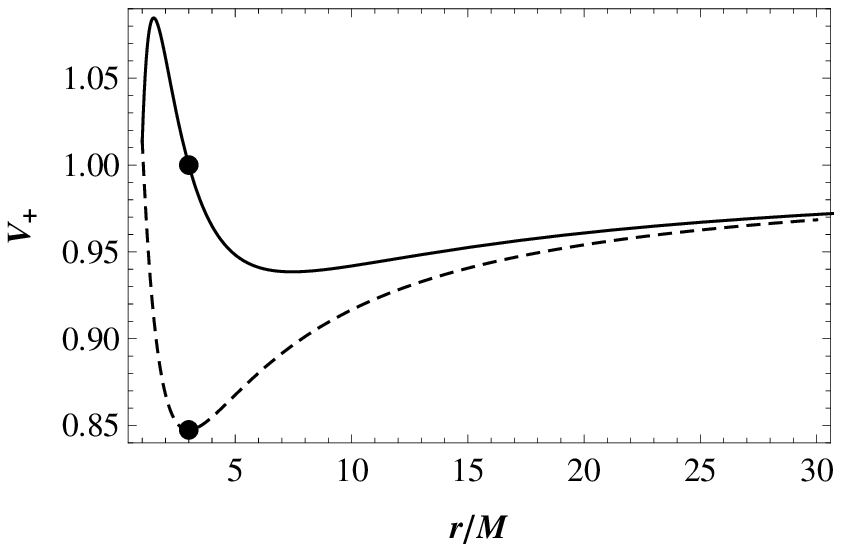}
\hspace{.2cm}\includegraphics[width=0.48\textwidth]{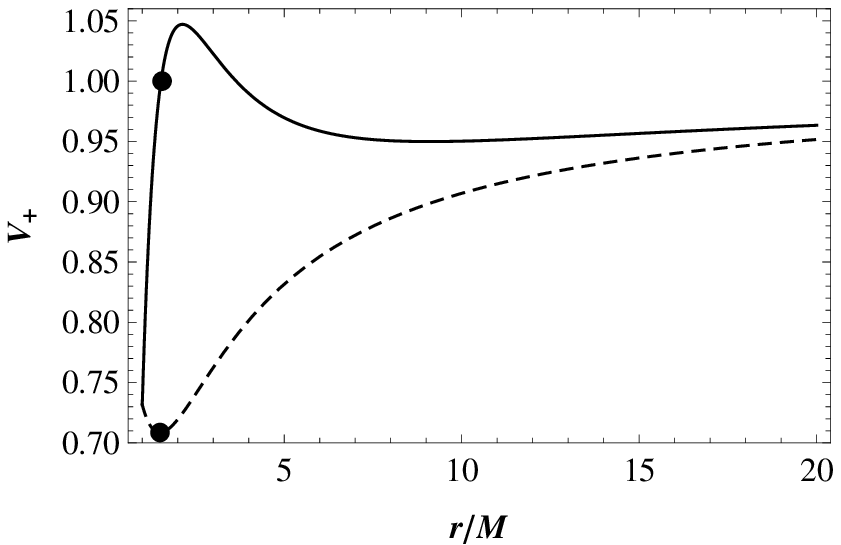}\non\\
&&\hspace{4.8cm}({\bf a})\hspace{7.8cm}({\bf b})\non
\ea
\caption{$V_+(r)$ for a particle before (dashed line) and after (solid line) getting kicked with $\dot{\theta}_k=\dot{\theta}_{\ce=1}$. (a) The particle kicked from the circular orbit at $r_o=3M$ accelerates away. (b) The particle kicked from the circular orbit at $r_o=3/2M$ accelerates toward the black hole. In both cases $a=M$.}\label{fig:epk}
\end{center}
\end{figure*}

The energetic freedom is not sufficient for the particle to escape when $a>0$, in general. Depending on the black hole's parameters and particle's initial conditions, the particle may accelerate both away or toward the black hole. $V_+(r)$ has only one maximum. The particle will therefore experience only one radial turning point. Hence, {\it the sign of the radial acceleration just after the kick $\ddot{r}(r_o)$ determines whether the particle escapes or gets captured}. Using Eq.~(\ref{epe}) we write an expression for $\ddot{r}(r)$ as
\begin{equation}
\ddot{r}(r)=-\frac{r^3+a^2(r+2M)}{2r^3}[\ce-V_-(r)] V'_+(r).
\end{equation}
Therefore, $\ddot{r}(r_o)\propto-V'_+(r_o)$ since $\ce>V_-(r_o)$. Figure~\ref{fig:epk} shows an example of capture and another of escape.

Careful analysis of $V'_+(r_o)$ reveals that there are {\it three} distinct regions in which the kicked particle accelerates in a specific way. The three regions are as follows:
\begin{itemize}

  \item {\bf Escape region}: For any $\dot{\theta}_k$ value $\ddot{r}(r_o)>0$ in this region. The acceleration is proportional to $|\dot{\theta}_k|$. The escape region is given by $r>r_{esc}$, where $r_{esc}$ is given by the equation
      \begin{equation}
      \hspace{12mm}(r_{esc}-3M)r_{esc}^2+a^2(r_{esc}+M)=0.
      \end{equation}
      The ISCO is located in this region when $a\lesssim0.853M$.

  \item {\bf Capture region}: In this region $\ddot{r}(r_o)<0$ for any $\dot{\theta}_k$. The stronger the kick, the faster the capture is. This region lies between $r_{cap}$ and the black hole's event horizon, where $r_{cap}$ is given by the equation
      \begin{eqnarray}
      &&\:\:M^{1/2}(ar_{cap}^2+a^3)+(r_{cap}-3M)r_{cap}^{5/2} \nonumber \\
      &&\hspace{2.6cm}+a^2(r_{cap}-M)r_{cap}^{1/2}=0.
      \end{eqnarray}
      The orbit at $r_o=M$ (when $a=M$), where Eq.~(\ref{ene0}) reduces to $\ce=1/\sqrt{3}$ for any $\dot{\theta}_k$, is an exception. The ISCO is located in the capture region for $M\geq a\gtrsim0.952M$.

  \item {\bf The critical escape region}: The particle acceleration is more involved in this region because its direction depends on $|\dot{\theta}_k|$ value. In particular, $\ddot{r}(r_o)>0$ if $|\dot{\theta}_k|$ is below some critical value $|\dot{\theta}_c|$. When $|\dot{\theta}_k|>|\dot{\theta}_c|$, the acceleration becomes inwards. For orbits with $|\dot{\theta}_{\ce=1}|>|\dot{\theta}_c|$, the particle can never escape. The critical escape region lies between the escape and capture regions. The critical kick angular velocity $\dot{\theta}_c$ is determined by
      \begin{equation}
       V'_+(r_o,\dot{\theta}_c)=0 ;\centering \hspace{5mm} r_{cap}<r_o<r_{esc}.
      \end{equation}

      In Fig.~\ref{fig:vc} we plot $|\dot{\theta}_c|$ and $|\dot{\theta}_{\ce=1}|$ vs. $r_o$ for $a=0.95M$. We see that $|\dot{\theta}_c|$ vanishes at $r_{cap}\approx1.92M$ and approaches infinity as $r_o$ approaches $r_{esc}\approx2.49M$. Figure~\ref{fig:rcvc} shows how the initial orbit radius $r_o$ at which $|\dot{\theta}_c|=|\dot{\theta}_{\ce=1}|$ changes with $a$. It is always greater than $r_{cap}$.
      \begin{figure}[!htb]
      \centering
      \includegraphics[width=.48\textwidth]{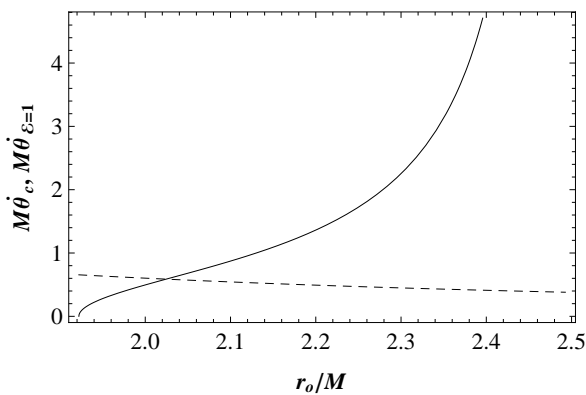}
      \caption{$|\dot{\theta}_c|$ (solid) and $|\dot{\theta}_{\ce=1}|$ (dashed) vs. $r_o$ for $a=0.95M$. $|\dot{\theta}_c|$ vanishes at $r_{cap}$ and approaches infinity as $r_o$ approaches $r_{esc}$.}
      \label{fig:vc}
      \end{figure}
      \begin{figure}[h!]
      \centering
      \includegraphics[width=0.48\textwidth]{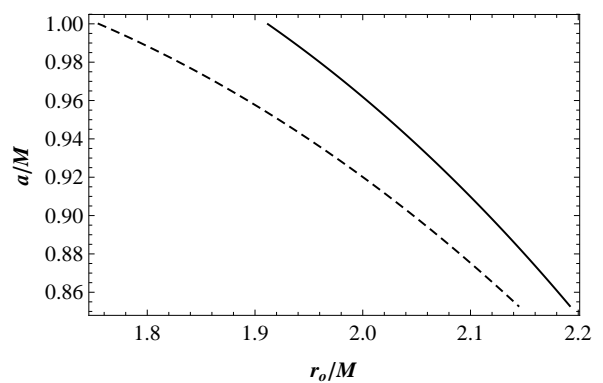}
      \caption{The radius of the initial orbit $r_o$ at which $|\dot{\theta}_c|=|\dot{\theta}_{\ce=1}|$ (solid) as a function of $a$. The dashed curve is $r_{cap}$.}
      \label{fig:rcvc}
      \end{figure}
\end{itemize}

\noindent Figure \ref{fig:reg} shows the three regions along with the ISCO and how they change with $a$. Incorporating all of the restrictions above, a particle in a circular orbit around a Kerr black hole kicked in the direction normal to the orbit can escape in the following two cases:
\begin{figure}[h!]
  \centering
  \includegraphics[width=0.48\textwidth]{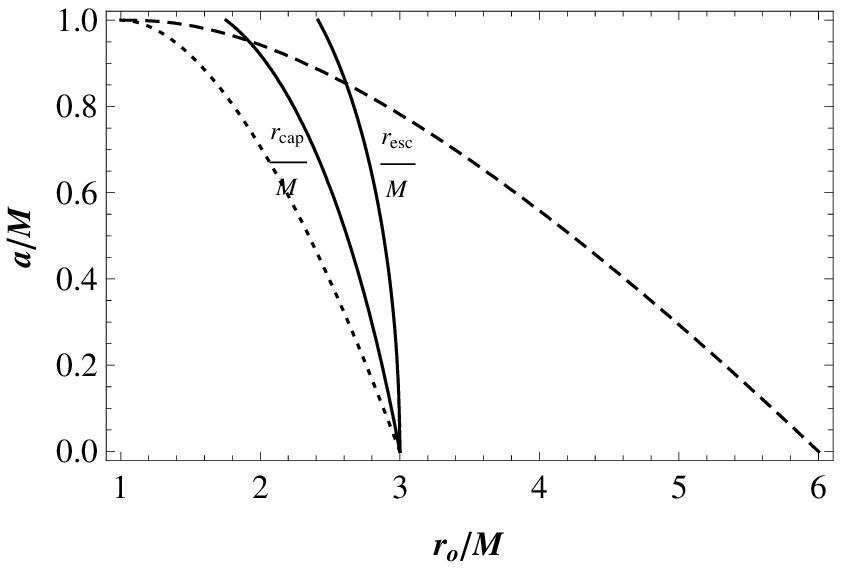}
  \caption{The dependence of $r_{esc}$, $r_{cap}$ and $r_{ms}$ (dashed) on $a$. The escape region is to the right of $r_{esc}$, the capture region is to the left of $r_{cap}$, while the critical escape region is the one in between. The dotted line is $r_{lc}$}
  \label{fig:reg}
\end{figure}
\begin{enumerate}
    \item Its initial orbit is in the escape region, $r_o\geq r_{esc}$.
    \item Its initial orbit is in the critical escape region, $r_{cap}<r_o<r_{esc}$, where it is possible to have $|\dot{\theta}_{\ce=1}|\leq |\dot{\theta}_k|<|\dot{\theta}_c|$.
\end{enumerate}

\section{Escape velocity of a Charged Particle} \label{s3}

\subsection{Weakly Magnetized Kerr Black Holes}

We follow the magnetization procedure introduced by Wald \cite{Wald}. In a Ricci flat spacetime a Killing vector $\xi^{\mu}$ obeys the equation
\begin{equation}
\xi^{\mu\:\:\:\: ;\nu}_{\:\:\;;\nu}=0.
\end{equation}

\noindent This is identical to the source-free Maxwell equations for a four-potential $A^{\mu}$ in the Lorentz gauge ($A^{\mu}_{\:\:\:;\mu}=0$),
\begin{equation}
A^{\mu\:\:\:\: ;\nu}_{\:\:\;;\nu}=0.
\end{equation}
Therefore, any linear combination of the Killing vectors the spacetime admits will serve as a solution to the Maxwell equations.

For the Kerr metric the choice
\begin{equation}\label{empot}
A^{\mu}=\frac{B}{2}\xi^{\mu}_{(\phi)}
\end{equation}

\noindent corresponds to an axisymmetric magnetic field that has strength $B$ asymptotically \cite{Wald,AG,AO}. It is this potential that will be used in this work.

The dynamics of a charged particle of mass $m$ and charge $q$ in an electromagnetic field in curved spacetime is governed by the equation
\begin{equation}\label{de}
m{u}^{\nu}\nabla_{\nu}u^{\mu}=qF^{\mu}_{\;\;\rho}u^{\rho},
\end{equation}

\noindent where $F_{\;\;\nu}^{\mu}$ is the electromagnetic field tensor given by
\begin{equation}
F_{\mu\nu}=A_{\nu,\mu}-A_{\mu,\nu}.
\end{equation}

\nin In the frame of an observer with four-velocity $u^\mu$, the electric and magnetic fields are, respectively
\begin{eqnarray}
E^{\mu}&=&F^{\mu\nu}u_{\nu}, \\
B^{\mu}&=&-\frac{1}{2}\frac{\varepsilon^{\mu\nu\lambda\sigma}}{\sqrt{-g}}F_{\lambda\sigma}u_{\nu},
\end{eqnarray}
where $g=\mbox{det}(g_{\mu\nu})$ and $\varepsilon_{0123}=+1$.

\nin The generalized four-momentum of the particle is
\begin{equation}
P^{\mu}=mu^{\mu}+qA^{\mu}.
\end{equation}

The weak field approximation breaks down when the magnetic field creates curvature comparable to that made by the black hole's mass, or
\begin{equation}
B^2\sim M^{-2}.
\end{equation}

\nin In conventional units, the Wald approximation fails when
\begin{equation}
B\sim \frac{k^{1/2}c^3}{G^{3/2}M},
\end{equation}

\nin where $k$ is the Coulomb constant. For a solar mass black hole one gets $B\sim10^{19}$Gauss. The typical magnetic field strength near a black hole's horizon has been estimated to be $\sim10^8$G ($10^{-15} \:\text{meter}^{-1}$) for stellar mass black holes and $\sim10^4$G ($10^{-19} \:\text{meter}^{-1}$) for supermassive black holes \cite{SGBPN,PSGN}. These estimates validate ignoring corrections to the metric due to the presence of the magnetic field. Despite that $B$ is "tiny" its effect on the dynamics is significant since $q/m=2.04\times 10^{21} (1.11\times 10^{18})$ for electrons (protons). For electrons (protons) near a typical stellar mass black hole ${qB}/{m}\sim10^7\:(10^3) \:\text{meter}^{-1}$ and near a typical supermassive black hole $qB/m\sim10^3\:(10^{-1}) \:\text{meter}^{-1}$.

\subsection{Circular Orbits}

The introduction of the magnetic field breaks down the Carter constant. It can be easily checked that
\begin{equation}
\dot{\cal K}\neq 0.
\end{equation}
The particle's energy and azimuthal angular momentum are constants of motion since the Lie derivatives of the electromagnetic potential~(\ref{empot}) with respect to the Killing vectors vanish
\begin{equation}
{\cal L}_{\xi^{\nu}_{(t)}}A^{\mu}={\cal L}_{\xi^{\nu}_{(\phi)}}A^{\mu}=0.
\end{equation}
The specific energy $\cal E$ and azimuthal angular momentum $\cal L$ are
\begin{eqnarray}
-{\cal E}&=&P_{\mu}\xi^{\mu}_{(t)}/m \non \\
&=&\left(\frac{2Mr}{\Sigma}-1\right)\dot{t}-\frac{2aMr}{\Sigma}(b+\dot{\phi})\sin^2{\theta}, \label{mage}\:\:\:\:\:\:\\
{\cal L}&=&P_{\mu}\xi^{\mu}_{(\phi)}/m \non \\
&=&\left[-\frac{2aMr}{\Sigma}\dot{t}+\frac{A}{\Sigma}(b+\dot{\phi})\right]\sin^2{\theta}, \label{magl}
\end{eqnarray}

\noindent where $b={qB}/{2m}$. Using these constants of motion and the normalization condition $u^{\mu}u_{\mu}=-1$ we write
\begin{eqnarray}
&&\dot{t}=\ce+\frac{2Mr[(r^2+a^2)\ce-a\cl]}{\Delta\Sigma},\label{tdeq}\\
&&\dot{\phi}=\frac{\cl}{\Sigma\sin^2\theta}+\frac{a(2Mr\ce-a\cl)}{\Delta\Sigma}-b\label{phideq}\\
&&\Sigma^2\left(\dot{r}^2+\Delta\dot{\theta}^2\right)=A\ce^2-4aM\ce\cl r-\Delta\Sigma(1-2b\cl) \nonumber\\
&&\hspace{2.4cm}+\frac{\cl^2(2Mr-\Sigma)}{\sin^2\theta}-b^2A\Delta\sin^2\theta.\label{ethd} \:\:\:\:\:\:\:\:
\end{eqnarray}
The $r$ and $\theta$ components of the dynamical equation (\ref{de}) are written in the appendix below.

Equations (\ref{tdeq})--(\ref{ethd}), (\ref{req}) and~(\ref{teq}) are invariant under reflection with respect to the equatorial plane~(\ref{sym1}). They are also invariant under the symmetry transformations
\begin{eqnarray}\label{sym2}
&&\phi\rightarrow-\phi, \hspace{3mm} \dot{\phi}\rightarrow-\dot{\phi}, \hspace{3mm} \cl\rightarrow-\cl, \non\\
 &&\hspace{10mm} a\rightarrow -a , \hspace{3mm} b\rightarrow -b.
\end{eqnarray}

\nin There are {\it four} dynamically distinct modes of motion. They are determined by the four combinations of the signs of $b\cl$ and $a\cl$. As before we fix $\cl$ to be positive. We just alter the signs of $a$ and $b$ to consider the four cases. We refer to the $b>0$ motion as anti-Larmor and to the $b<0$ motion as Larmor. For circular orbits, the radial acceleration of the particle $f^1=\frac{q}{m}(F^1_{\;\;0}\dot{t}+F^1_{\;\;3}\dot{\phi})$ is positive for the anti-Larmor motion and negative for Larmor motion.

Equation (\ref{ethd}) simplifies in the equatorial plane to
\begin{eqnarray}\label{rpm}
r^3\dot{r}^2&=&(\ce^2-b^2\Delta)[r(r^2+a^2)+2Ma^2]- \nonumber\\
&&4aM\ce\cl-r\Delta(1-2b\cl)-\cl^2(r-2M).\:\:\:\:\:\:\:\:
\end{eqnarray}

\nin Let us define the positive semi-definite function ${\cal R}(r)$ to be the right hand side of Eq.~(\ref{rpm}):
\begin{eqnarray}
&&{\cal R}(r):=(\ce^2-b^2\Delta)[r(r^2+a^2)+2Ma^2]- \nonumber\\
&&\hspace{10mm}4aM\ce\cl-r\Delta(1-2b\cl)-\cl^2(r-2M). \:\:\:\:\:\:\:\:
\end{eqnarray}

\nin Then using the circular orbit conditions ${\cal R}(r)=0$ and ${\cal R}'(r)=0$ one obtains, respectively,
\begin{eqnarray}
&&(\ce^2-b^2\Delta)[r(r^2+a^2)+2Ma^2]-4aM\ce\cl \non\\
&&\hspace{12mm}-r\Delta(1-2b\cl)-\cl^2(r-2M)=0,\:\:\:\label{co1}
\end{eqnarray}
and
\begin{eqnarray}
&&2b^2(r-M)[r(r^2+a^2)+2Ma^2]+ \non \\
&&(1-2b\cl)[2r(r-M)+\Delta]+\cl^2- \non \\
&&\hspace{15.5mm}(\ce^2-b^2\Delta)(3r^2+a^2)=0.\label{co2}
\end{eqnarray}

\nin The extra condition for ISCOs ${\cal R}''(r)=0$ gives
\begin{eqnarray}
&&(2b\cl-1)(3r-2M)+3\ce^2r\non \\
&&\hspace{10mm}-2b^2r[r(5r-6M)+3a^2]=0.
\end{eqnarray}

\nin It is very difficult to solve Eqs.~(\ref{co1}) and~(\ref{co2}) to obtain analytic expressions for $\ce_o$ and $\cl_o$. Instead, we solve these equations numerically. We also require that $\dot{t}>0$ to exclude past-directed solutions.

\begin{figure*}[ht]
\begin{center}
\ba
&&\hspace{0cm}\includegraphics[width=0.48\textwidth]{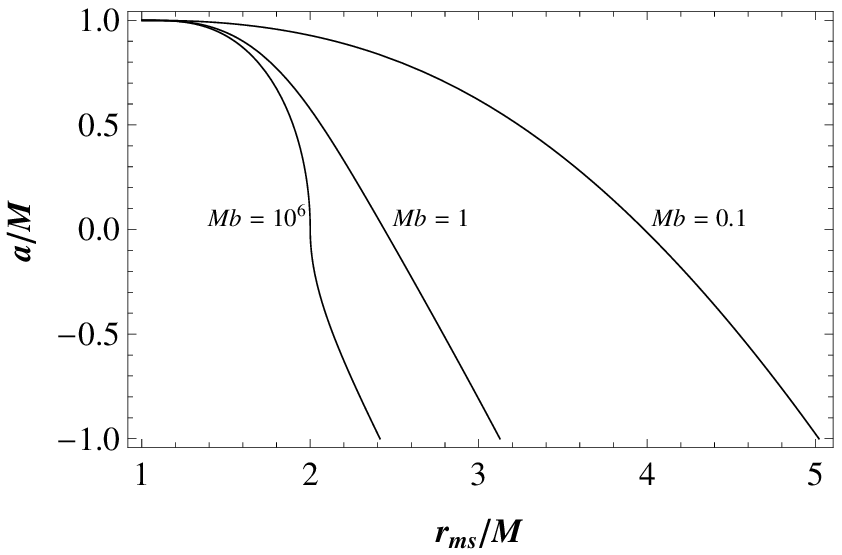}
\hspace{.2cm}\includegraphics[width=0.48\textwidth]{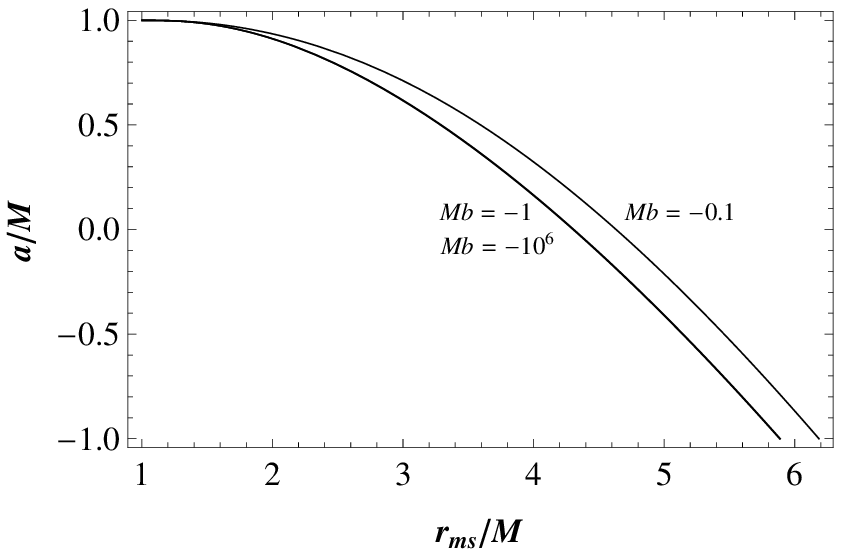}\non\\
&&\hspace{4.8cm}({\bf a})\hspace{7.8cm}({\bf b})\non
\ea
\caption{The ISCO's radius $r_{ms}$ dependence on $a$ for different values of the magnetic parameter $b$ for (a) anti-Larmor motion and (b) Larmor motion.}\label{rmsb}
\end{center}
\end{figure*}

\begin{figure*}[ht]
\begin{center}
\ba
&&\hspace{0cm}\includegraphics[width=0.48\textwidth]{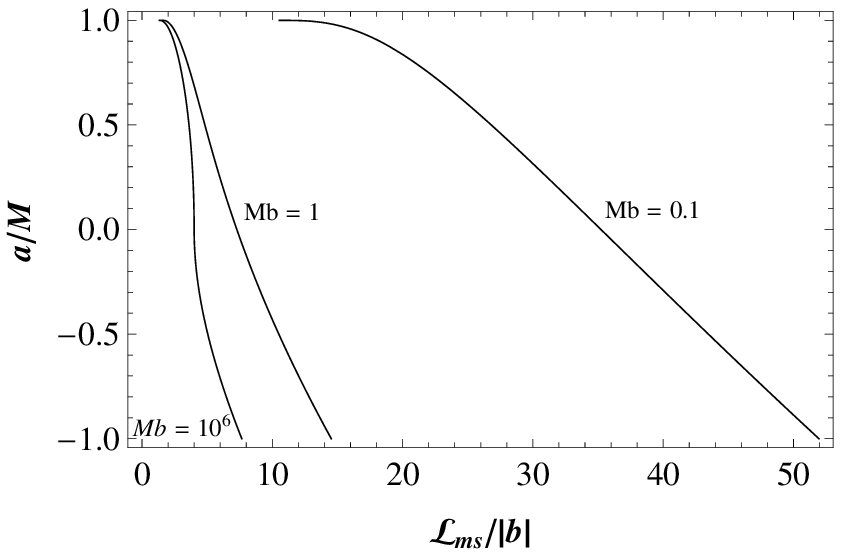}
\hspace{.2cm}\includegraphics[width=0.48\textwidth]{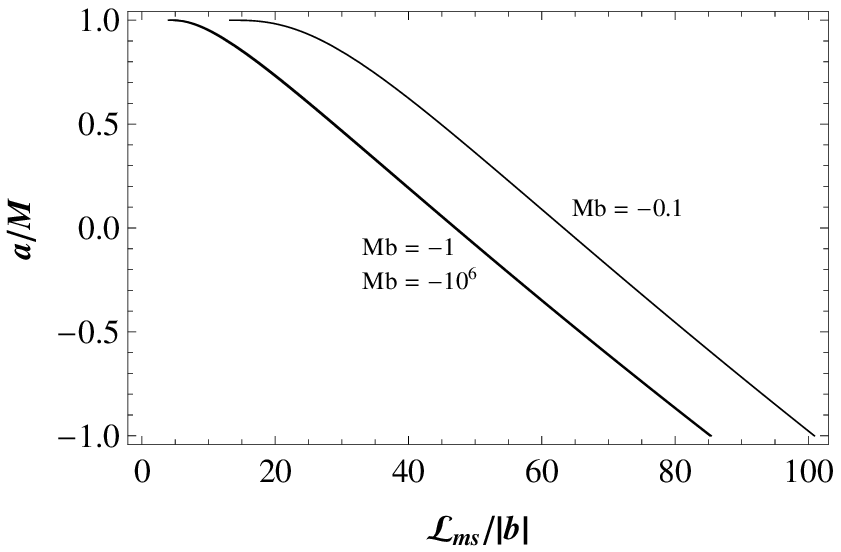}\non\\
&&\hspace{4.8cm}({\bf a})\hspace{7.8cm}({\bf b})\non
\ea
\caption{The ISCO's azimuthal angular momentum $\cl_{ms}$ dependence on $a$ for different values of the magnetic parameter $b$ for (a) anti-Larmor motion and (b) Larmor motion.}\label{Lmsb}
\end{center}
\end{figure*}

\begin{figure*}[ht]
\begin{center}
\ba
&&\hspace{0cm}\includegraphics[width=0.48\textwidth]{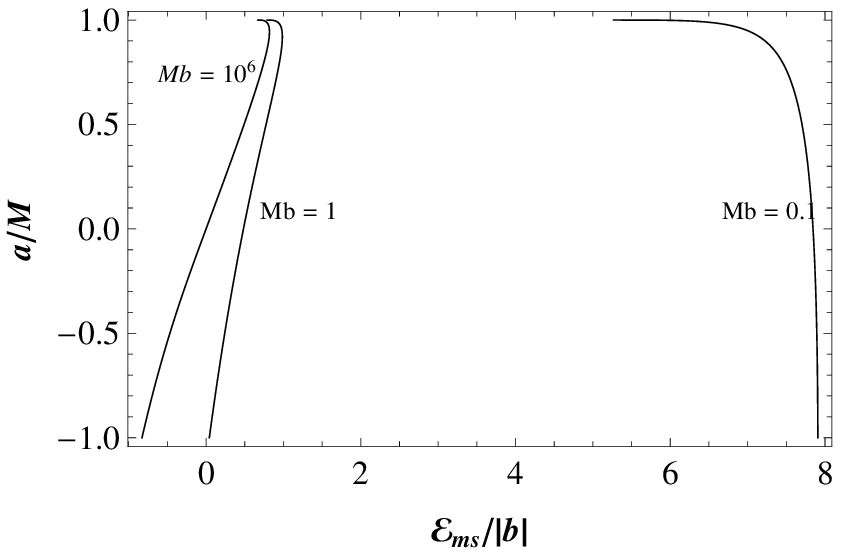}
\hspace{.2cm}\includegraphics[width=0.48\textwidth]{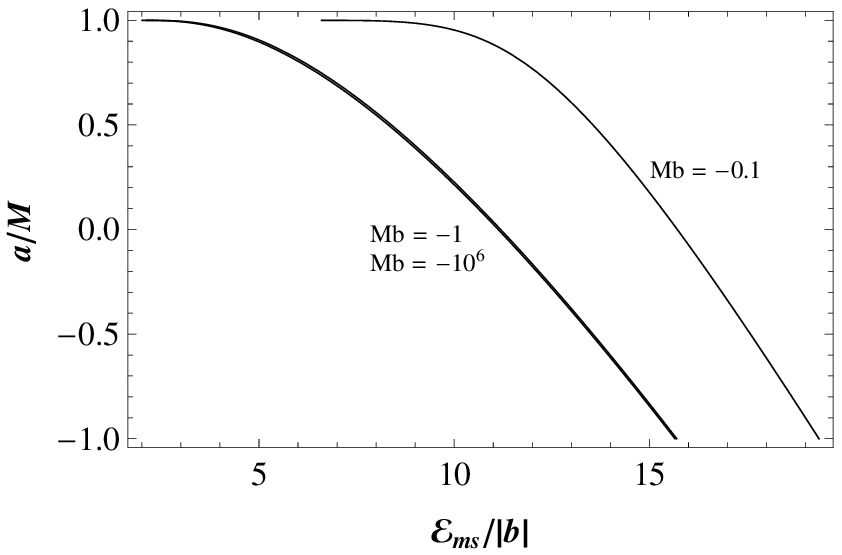}\non\\
&&\hspace{4.8cm}({\bf a})\hspace{7.8cm}({\bf b})\non
\ea
\caption{The ISCO's energy $\ce_{ms}$ dependence on $a$ for different values of the magnetic parameter $b$ for (a) anti-Larmor motion and (b) Larmor motion.}\label{Emsb}
\end{center}
\end{figure*}

It is interesting to see how the ISCO radius depends on $a$ for selected values of the magnetic parameter $b$. Knowing the dependence of the ISCO radius on $a$ is essential for measuring the spin of astrophysical black holes \cite{Bren}. The $a$--$r_{ms}$ curves for selected $b$ values are shown in Fig.~\ref{rmsb}. When $b=0$ Fig.~\ref{fig:arms} is reproduced. In both Larmor and anti-Larmor motions $r_{ms}$ gets closer to the black hole as $|b|$ increases. It converges to an asymptotic value as $|b|$ becomes large. The shift in $r_{ms}$ is more evident in the anti-Larmor motion. The value of $r_{ms}$ is different from the asymptotic values by less than $0.1\%$ when $5.8\times10^3M^{-1}<b<-0.82 M^{-1}$. For retrograde motion $r_{ms}$ is always outside the static limit. Figures~\ref{Lmsb} and~\ref{Emsb} show $\cl_{ms}$ and $\ce_{ms}$ corresponding to the ISCOs shown in Fig.~\ref{rmsb}. It is interesting that negative energy stable circular orbits can exist in the retrograde anti-Larmor motion. The possibility for the existence of negative energy states due to magnetic fields was pointed out in Ref.~\cite{PrDa} and further explored in Ref.~\cite{DhDa1}. The related energy-emission processes were discussed in Refs.~\cite{DhDa2,PWHD}. At $a=-M$, $\ce_{ms}$ becomes zero when $b=b_c$, where $Mb_c$ is the positive real root of
\begin{eqnarray}
&&45056x^{12}-52224x^{10}+3072x^8\non\\
&-&3776x^6+4656x^4-1320x^2=25.
\end{eqnarray}
Numerically, $x\approx 1.0534$. As $b$ increases further $\ce_{ms}$ becomes negative for a larger interval of $a>-M$. Asymptotically, $\ce_{ms}$ becomes negative for all retrograde anti-Larmor orbits and approaches a minimum of $2(1-\sqrt{2})Mb$ at $a=-M$ where $r_{ms}=(1+\sqrt{2})M$. This immense binding energy is intriguing. A charged particle of mass $m_q$ and $b\gg M^{-1}$ ending up in this 'superbound' state can give off energy
\begin{equation}
E=m_q\ce_{ms}=(\sqrt{2}-1)qBM.
\end{equation}
For typical stellar mass and supermassive black holes of masses $M_{St}$ and $M_{Su}$, respectively, this amounts to
\begin{eqnarray}
E=1.832\times10^6 (\frac{M_{St}}{M_{\odot}})\: \mbox{GeV}, \\
E=1.832\times10^2 (\frac{M_{Su}}{M_{\odot}})\: \mbox{GeV},\label{smbhe}
\end{eqnarray}
where $M_{\odot}$ is the solar mass. For a supermassive black hole of mass $M=10^{9.5}M_{\odot}$, Eq.~(\ref{smbhe}) gives $E\sim100$ Joules.

It should be noted that the correspondence between $r_{ms}$ and $a$ is {\it one-to-one} in all cases, after past-directed orbits are excluded. The equation for $r_{ms}$ given in Ref.~\cite{AO} yields future-directed solutions only when $b<b_c$.


\subsection{Three-Dimensional motion and conditions for escape from a circular orbit}

\noindent It does not seem possible to determine the escape conditions analytically since the equations of motion are non-integrable in general. Equations~(\ref{req}) and~(\ref{teq}) were solved numerically using the built-in {\it MATHEMATICA} 7.0 function NDSOLVE. We used the constant of motion $\ce$ as a gauge of error in the numerical solver. The deviation in $\ce$ is $\sim10^{-6}$ or less. Sometimes the error grows to $\sim10^{-3}$ when the integration time is very long. We can increase the accuracy of the solver to achieve much better accuracy. This is not a problem when few trajectories are plotted, but it is very time-consuming when the basins of attraction are generated (see below). That is because in generating them the equations of motions are integrated $\sim10^6$ times and we are concerned about the final state of the particle which is practically not modified by increasing the accuracy.

\begin{figure*}[ht]
    \begin{center}
    \ba
    &&\hspace{1cm}\includegraphics[width=0.25\textwidth]{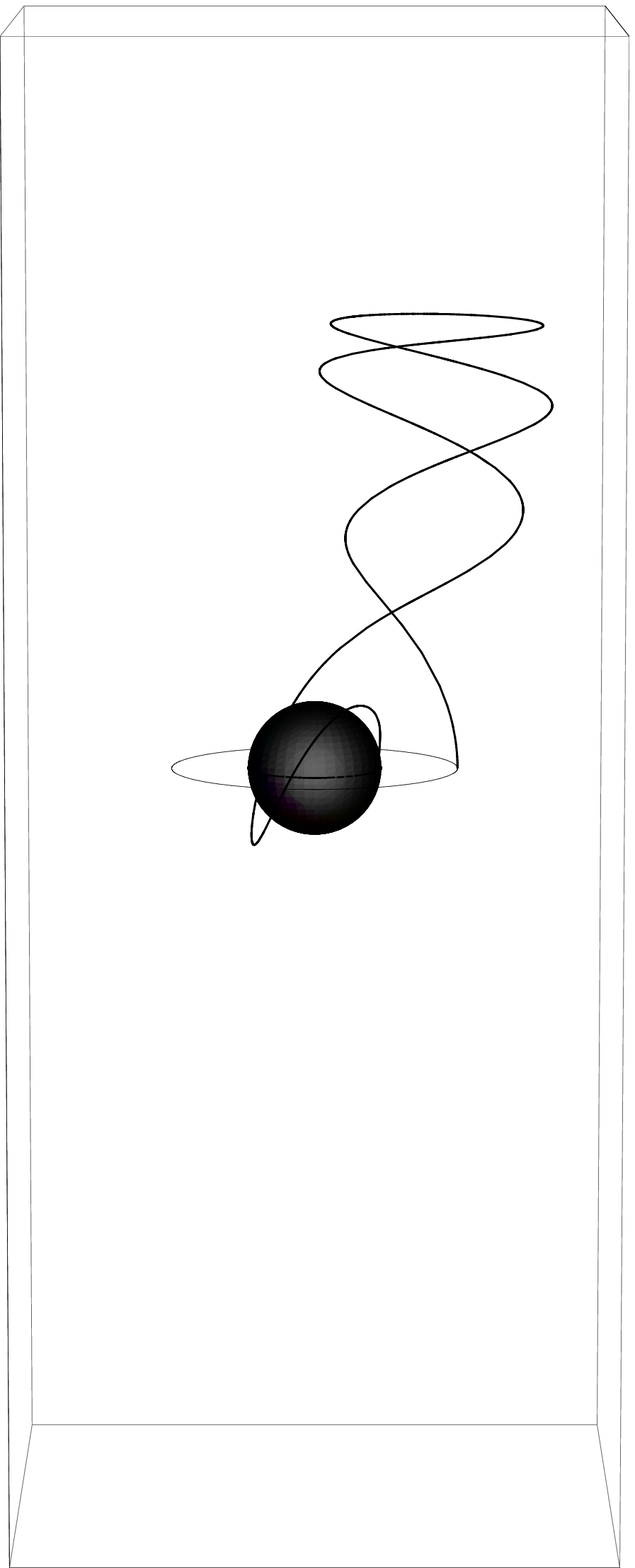}
    \hspace{1cm}\includegraphics[width=0.25\textwidth]{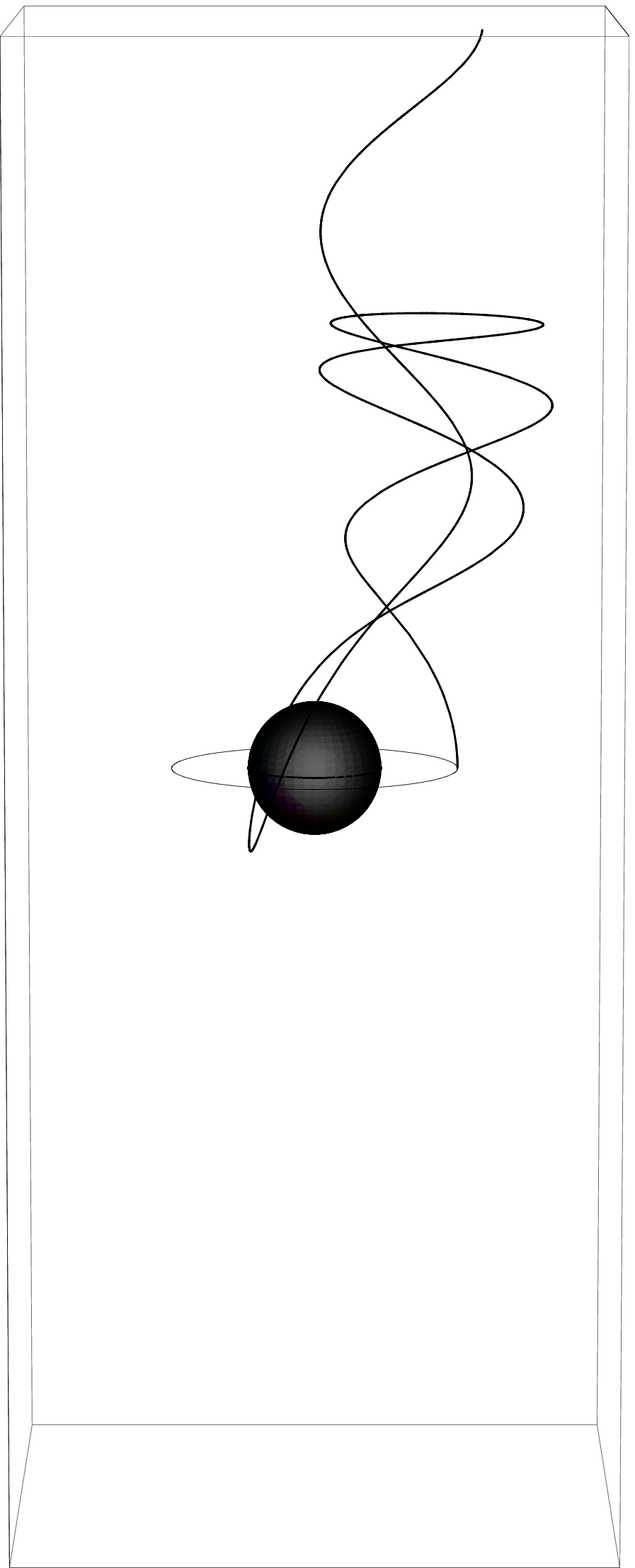}
    \hspace{1cm}\includegraphics[width=0.25\textwidth]{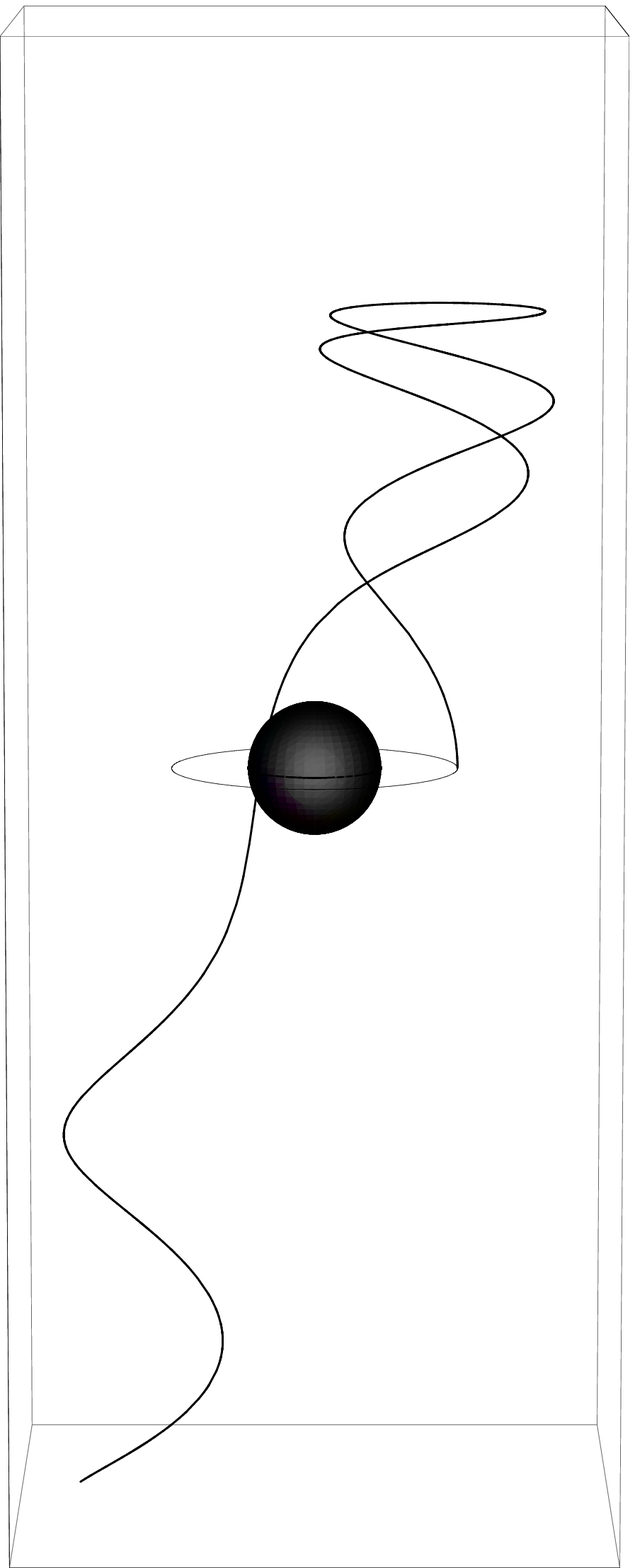}\non\\
    &&\hspace{3.cm}({\bf a})\hspace{4.4cm}({\bf b})\hspace{4.4cm} ({\bf c})\non
    \ea
    \caption{The trajectories of a charged particle initially at $r_o=4M$ kicked to three different energies (a): $\ce=1.0890$ (b): $\ce=1.0893$ (c): $\ce=1.0900$. In all cases $a=0.5M$ and $b=0.1 M^{-1}$. The particle is scattered to a different final state in every case.}\label{fig:tra}
    \end{center}
\end{figure*}

The numerical integration reveals that the escape and capture regions are more involved than those in the neutral particle case. In Fig.~\ref{fig:tra} the trajectories of a charged particle kicked up to three different energies  $\ce=1.0890$, $\ce=1.0893$ and $\ce=1.0900$ are shown. In this section we use $\ce$ to quantify the kick instead of $\dot{\theta}_k$, for convenience. The two are related by Eq.~(\ref{ethd}). The particle in each case ends up following a completely different trajectory despite the tiny difference between the energies. This extreme sensitivity to initial conditions is a characteristic of non-integrable and chaotic systems. To obtain a comprehensive view of the problem we need to identify which initial conditions lead to escape and which lead to capture.

In general the particle is fated to be captured by the black hole, escape it up (down) and approach $z=\infty$ ($z=-\infty$) asymptotically, or end up in an orbit 'meta-staple' within the computation time. Keeping the possible meta-stable orbits aside, the system therefore has three {\it attractors}. We need to use a method well-suited for analysing non-compact chaotic scattering systems.

An attractor of a dynamical system is a subset of the set of all possible states of the system which an orbit with certain initial conditions approaches asymptotically. The set of initial conditions which leads to an attractor is its {\it basin of attraction}.

The boundary between different basins of attraction in the space of initial conditions is a simple smooth curve (surface) in case of regular systems. In chaotic systems the basin-boundary is a {\it fractal} boundary. A fractal is a geometrical object that has fractal dimension $D_f$ larger than its topological dimension. A characteristic of fractals is the appearance of self-similar patterns persistent at any magnification.

Let us see how the basin of attraction plot looks like for a neutral particle first. We use the following color notation for all basins of attraction in this paper: (1) Green for escape to $z\rightarrow +\infty$ (2) Yellow for escape to $z\rightarrow -\infty$, (3) Red for capture and (4) Blue for meta-stable orbits.

\begin{figure}[h!]
  \centering
  \includegraphics[width=0.48\textwidth]{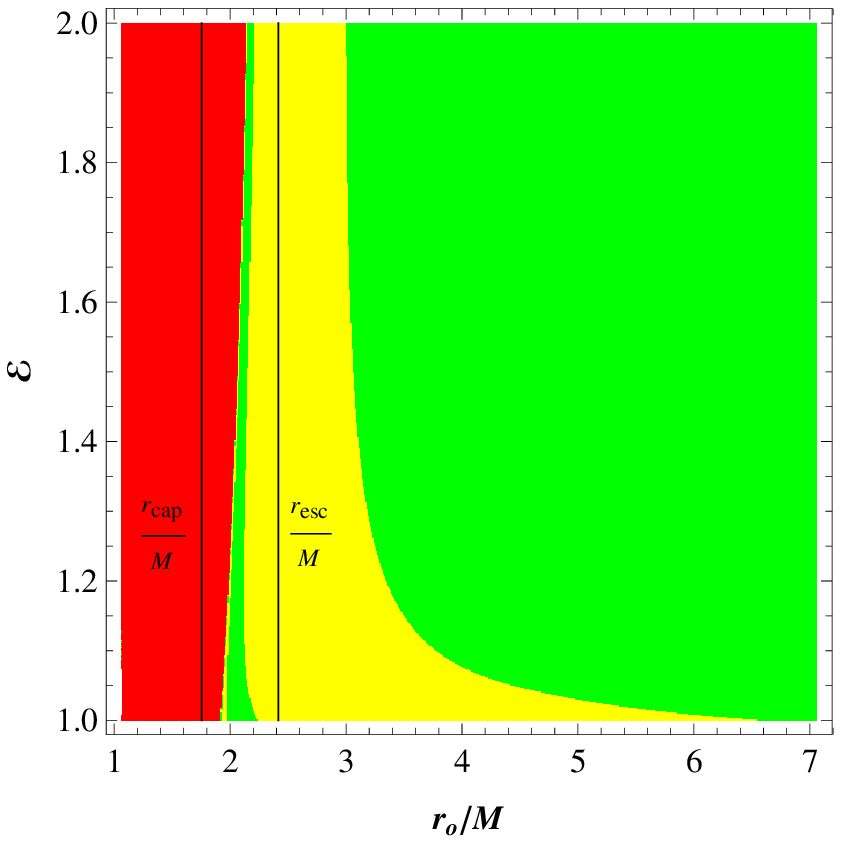}
  \caption{The basin of attraction plot for a neutral particle when $a=0.999M$. $r_{esc}$ and $r_{cap}$ are shown as well.}
  \label{fig:nb}
\end{figure}

Figure~\ref{fig:nb} shows the basin of attraction plot for a neutral particle generated numerically with initial values of $r_o\in[r_{lc},r_{lc}+6M]$ plotted horizontally and initial values of $\ce\in[1.0,2.0]$ plotted vertically. The resolution of the plot is 600$\times$600. We tackled the $a=0.999M$ case, where $r_{lc}=1.052M$, because the structure of the basin of attraction plot in this case is the richest. The basin boundaries are regular lines as they should be for a regular system. The structure of the escape and capture attractors is in accord with that described analytically in Sec.~\ref{s2}. The red color approaches $r_{esc}$ as $\ce$ becomes very large. The particle is backscattered near $r_{esc}$ and at low energies, where it barely makes it to escape.

\begin{figure*}[ht]
\begin{center}
\ba
&&\hspace{0.2cm}\includegraphics[width=0.3\textwidth]{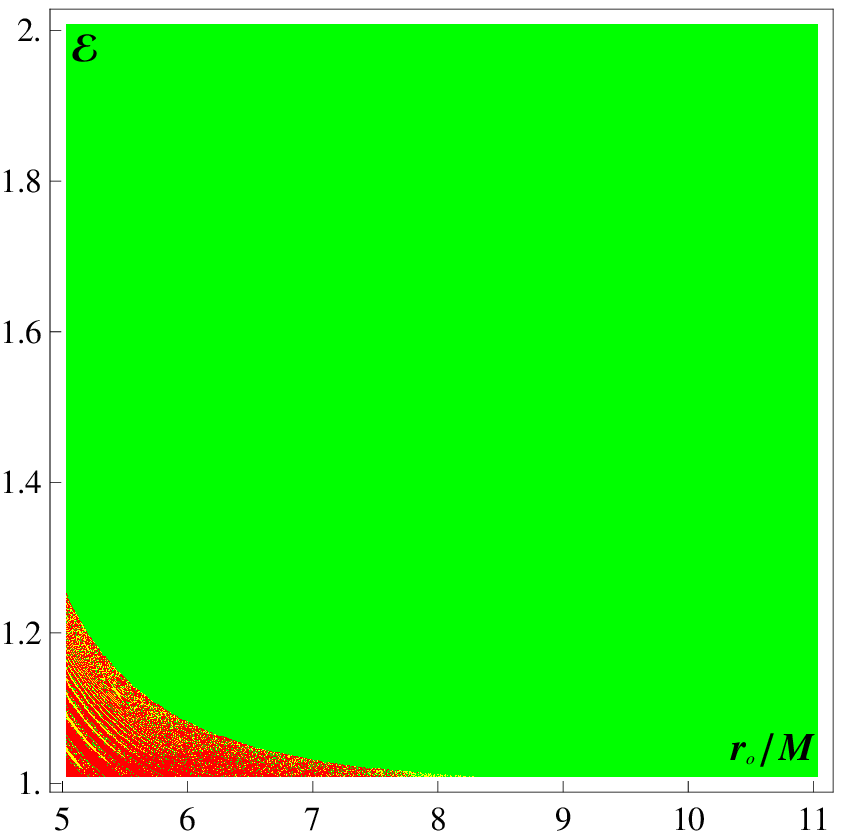}
\hspace{.2cm}\includegraphics[width=0.3\textwidth]{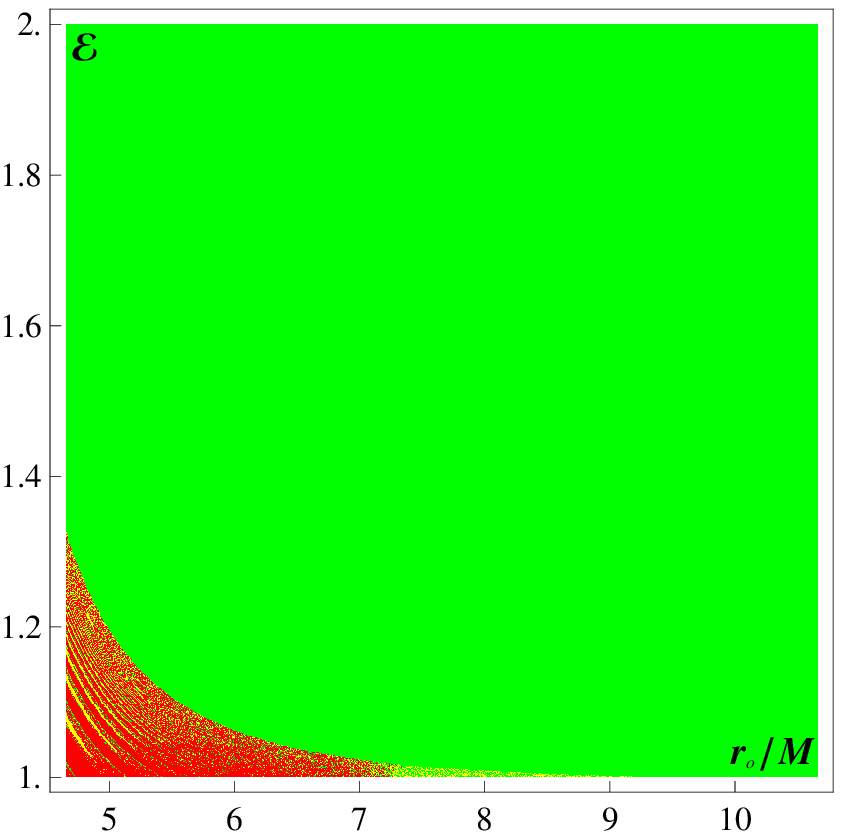}
\hspace{.2cm}\includegraphics[width=0.3\textwidth]{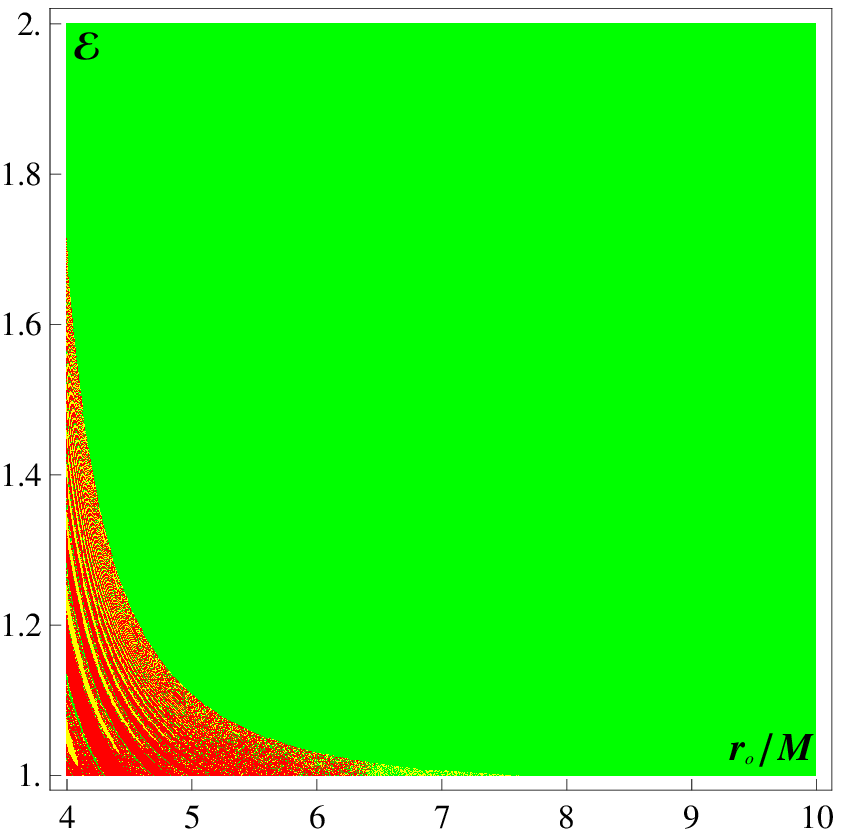}\non\\
&&\hspace{1.5cm}({\bf a}):\: a=-0.999M. \hspace{2.3cm}({\bf b}):\: a=-0.6M.\hspace{2.5cm} ({\bf c}):\: a=0.\non\\
&&\hspace{.2cm}\includegraphics[width=0.3\textwidth]{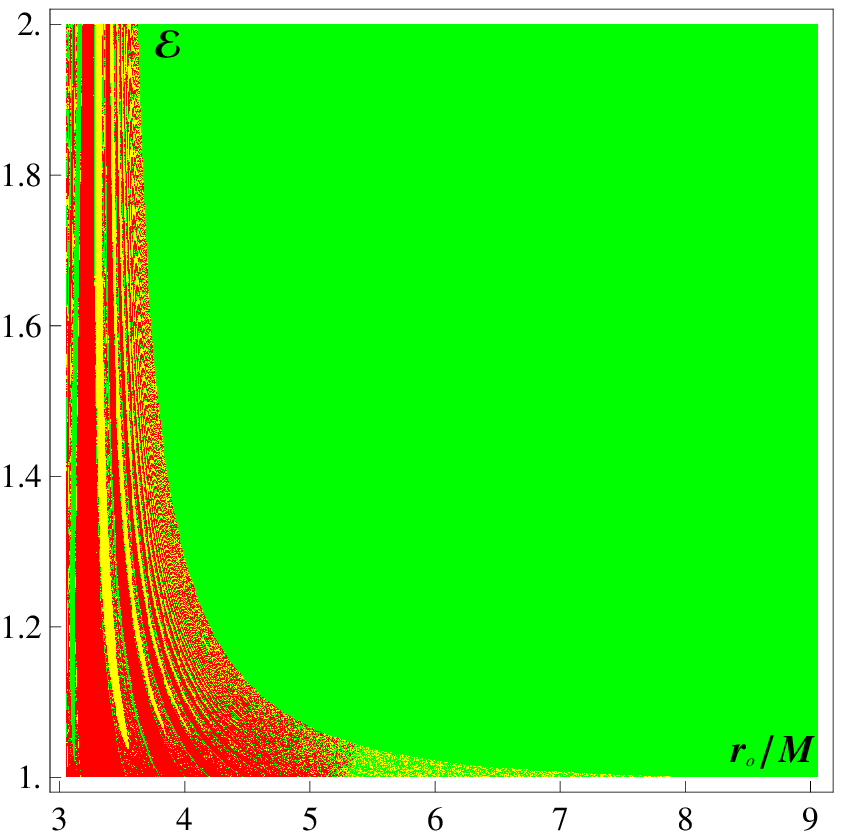}
\hspace{.2cm}\includegraphics[width=0.3\textwidth]{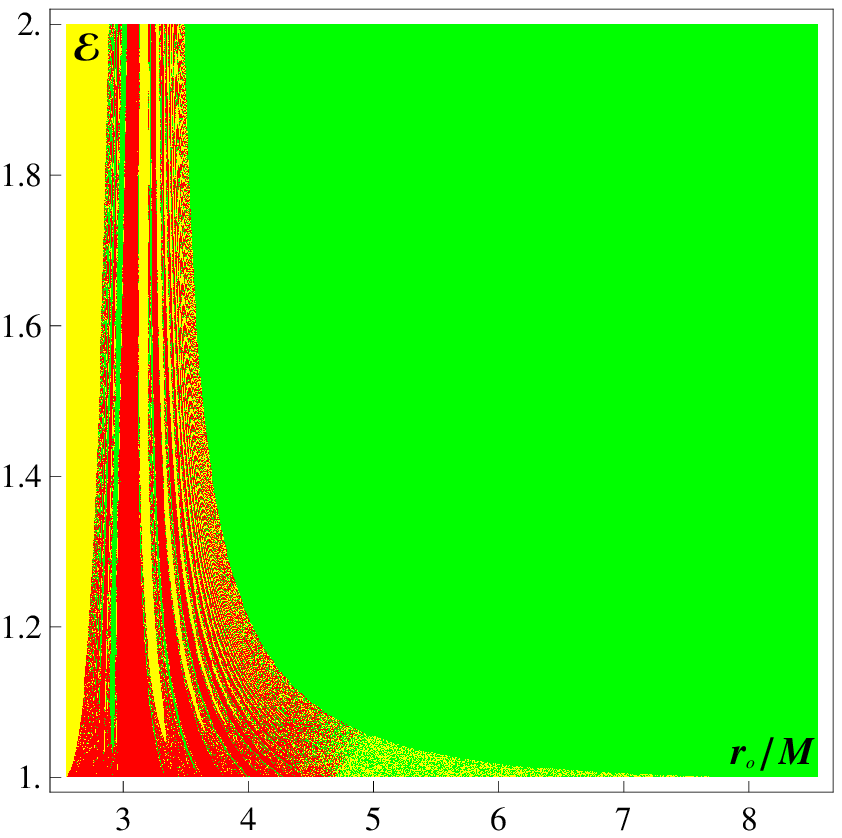}
\hspace{.2cm}\includegraphics[width=0.3\textwidth]{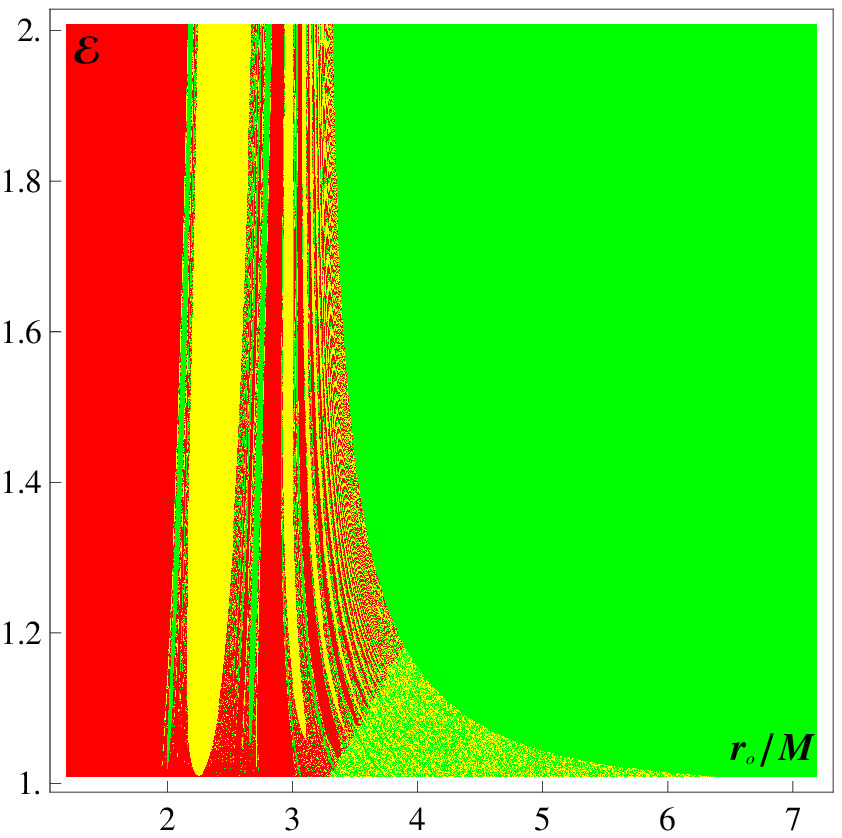}\non\\
&&\hspace{1.5cm}({\bf d}):\: a=0.6M. \hspace{2.9cm}({\bf e}):\: a=0.8M.\hspace{2.7cm} ({\bf f}):\: a=0.999M.\non
\ea
\caption{The basin of attraction plots for a charged particle with $b=0.1 M^{-1}$.}\label{fig:anti-lar}
\end{center}
\end{figure*}

\begin{figure*}[ht]
\begin{center}
\ba
&&\hspace{0.2cm}\includegraphics[width=0.3\textwidth]{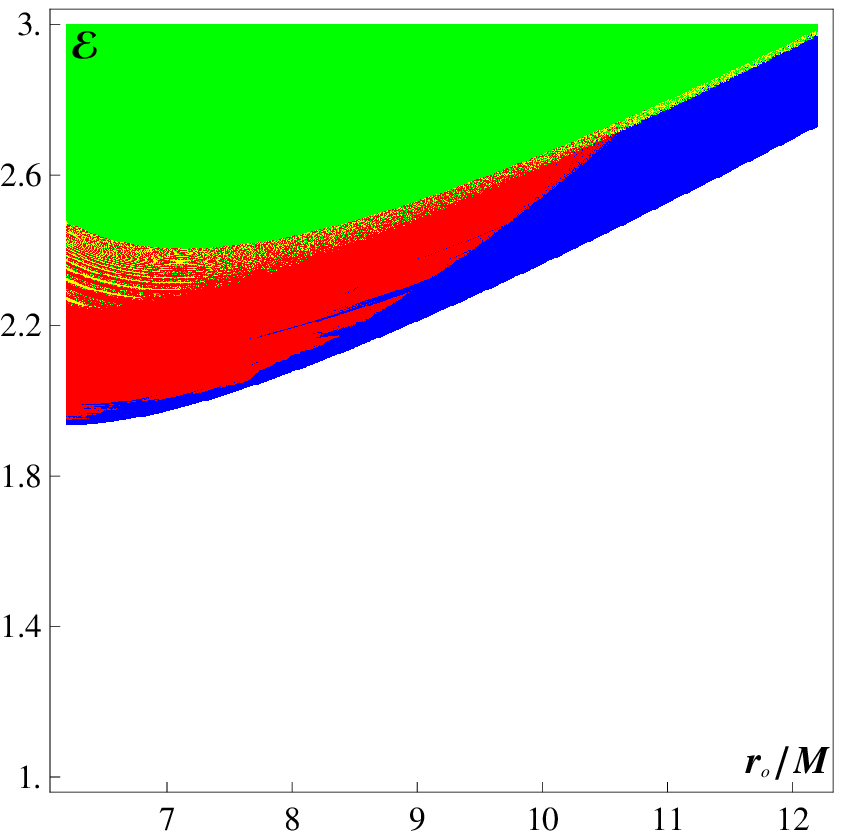}
\hspace{.2cm}\includegraphics[width=0.3\textwidth]{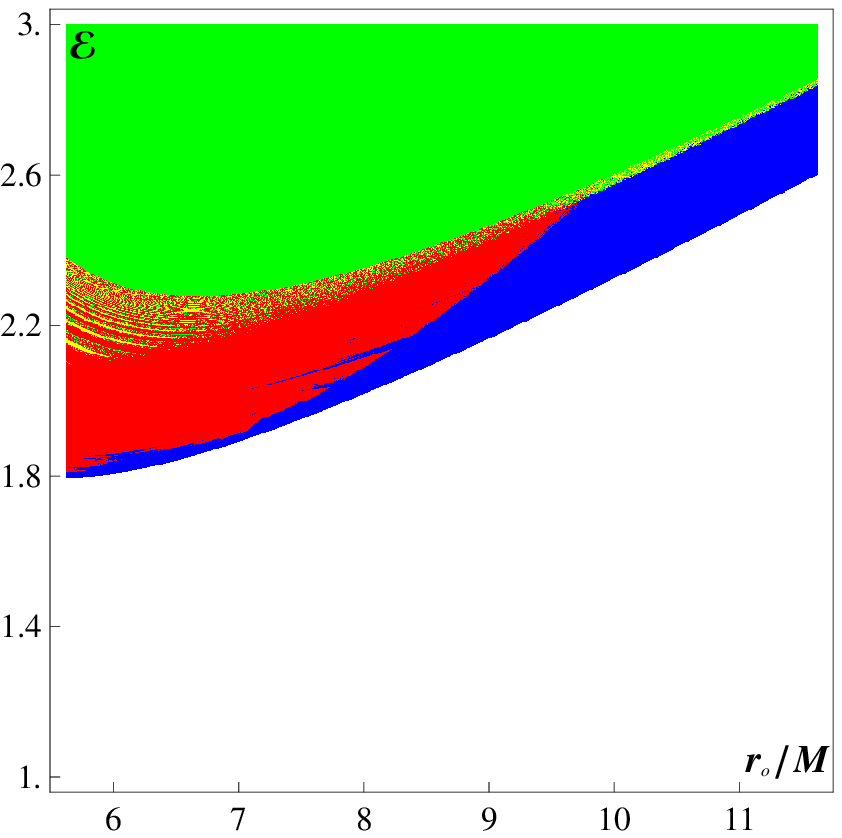}
\hspace{.2cm}\includegraphics[width=0.3\textwidth]{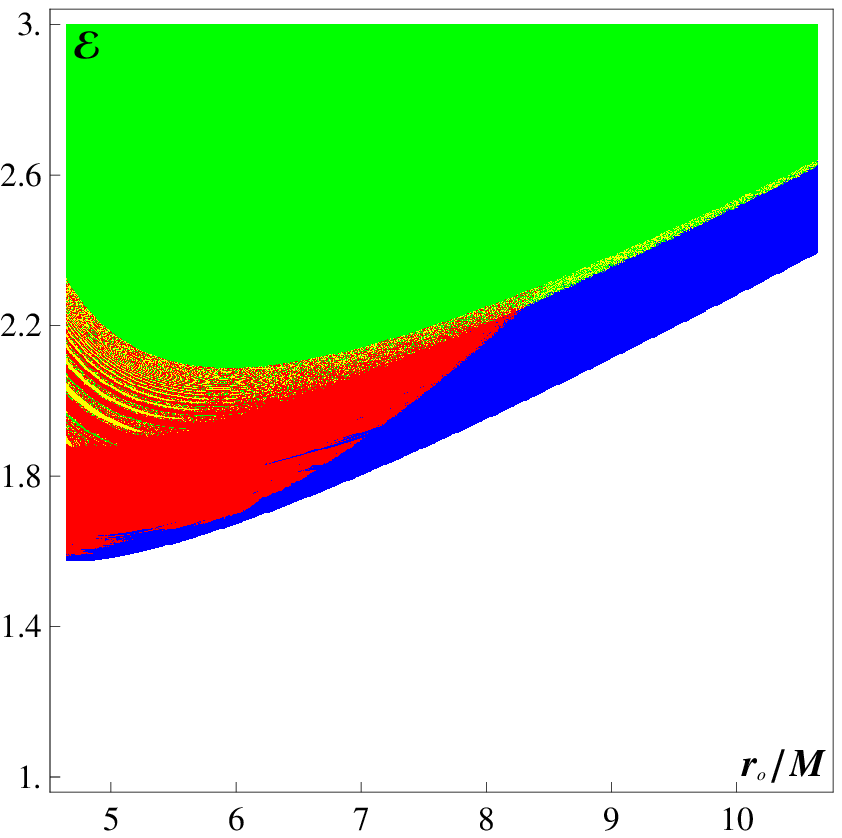}\non\\
&&\hspace{1.5cm}({\bf a}):\: a=-0.999M. \hspace{2.3cm}({\bf b}):\: a=-0.6M.\hspace{2.5cm} ({\bf c}):\: a=0.\non\\
&&\hspace{.2cm}\includegraphics[width=0.3\textwidth]{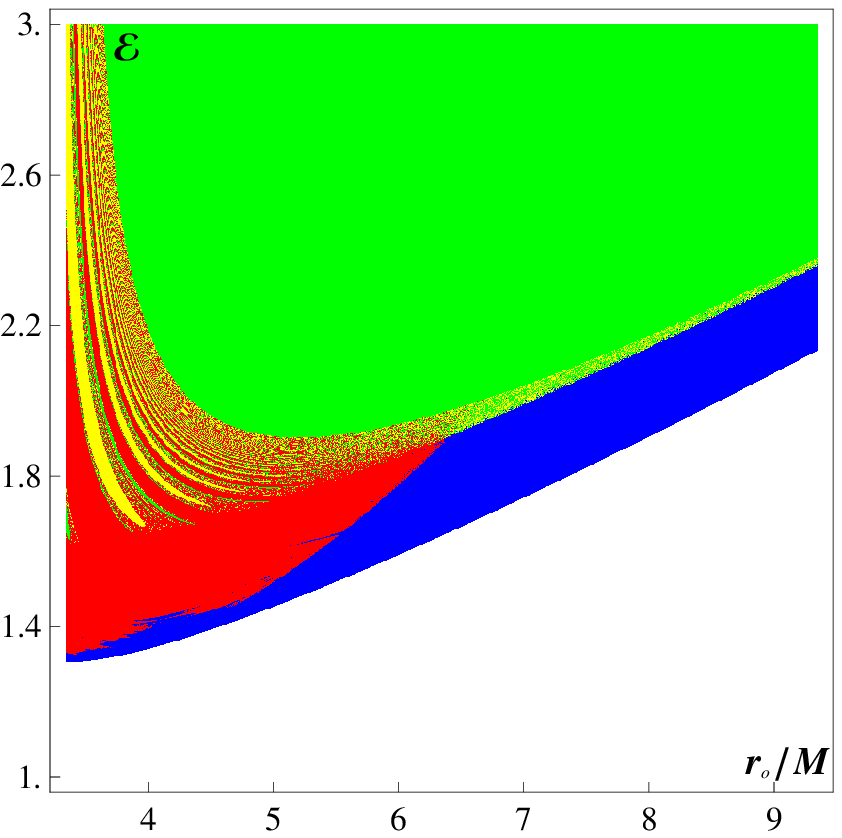}
\hspace{.2cm}\includegraphics[width=0.3\textwidth]{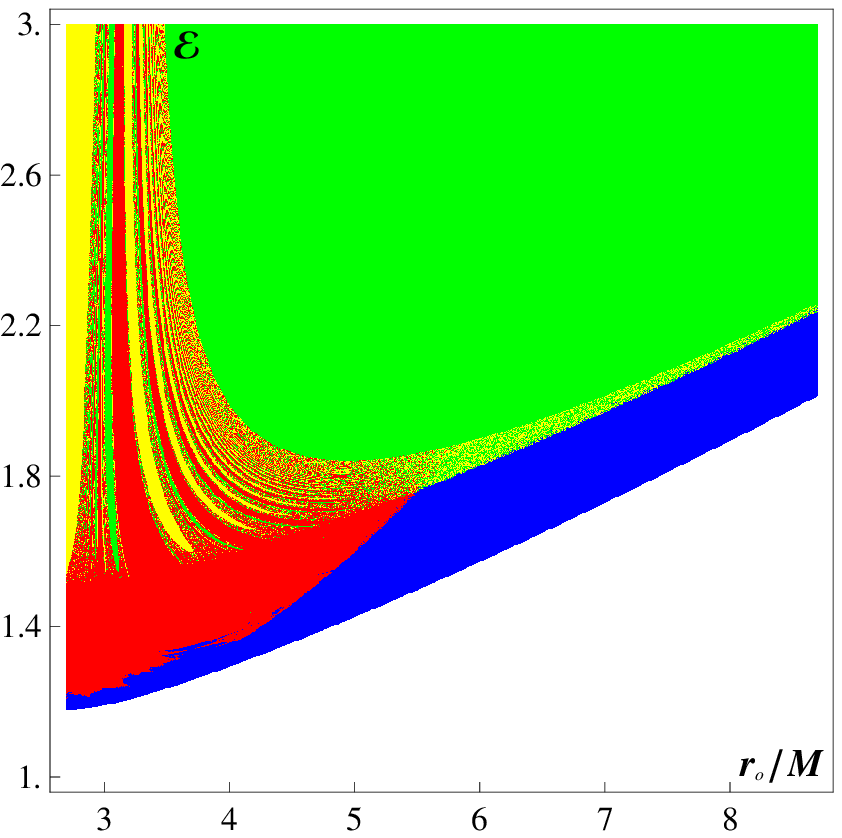}
\hspace{.2cm}\includegraphics[width=0.3\textwidth]{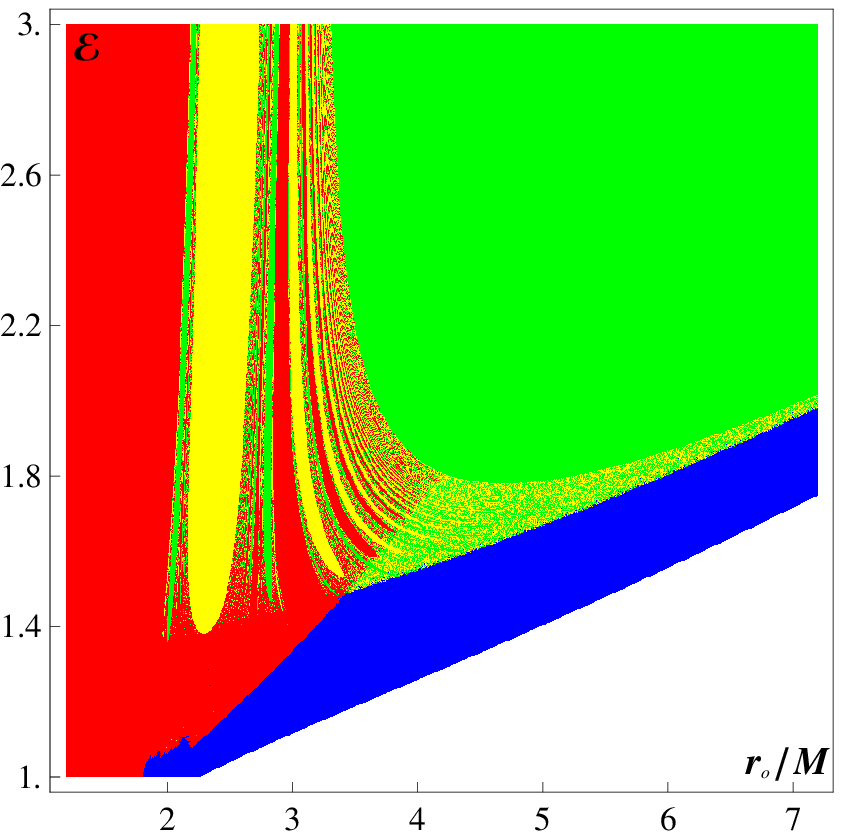}\non\\
&&\hspace{1.5cm}({\bf d}):\: a=0.6M. \hspace{2.9cm}({\bf e}):\: a=0.8M.\hspace{2.7cm} ({\bf f}):\: a=0.999M.\non
\ea
\caption{The basin of attraction plots for a charged particle with $b=-0.1 M^{-1}$.}\label{fig:lar}
\end{center}
\end{figure*}

Now we return to charged particles. Figure.~\ref{fig:anti-lar} shows the basin of attraction plots for anti-Larmor motion ($b=0.1 M^{-1}$) with initial values of $r_o\in[r_{ms},r_{ms}+6M]$ plotted horizontally and initial values of $\ce\in[1.0,2.0]$ plotted vertically. Figure~\ref{fig:lar} shows the Larmor motion ($b=-0.1 M^{-1}$) basin of attraction plots but with initial values of $\ce\in[1.0,3.0]$ since $\ce_o$ is usually considerably larger than 1. The white regions in the figures represent the energetically forbidden orbits. The value of $|b|$ considered here may be small compared to typical astrophysical values. Nonetheless, we find it appropriate to demonstrate the various aspects of the problem. The spin parameter $a$ was taken at selected values between $-1$ and $1$.

The state of the particle is considered an escape if it reaches $z=200M$. At this distance the gravitational potential can be well approximated by the Newtonian value of $-M/r\approx -M/z$. In cases for which $\dot{z}^2<2M/z$ the particle will return back and all three outcome are possible. This is the case with about $1\%$ of escape cases, especially when $\ce$ is just above $1$. The maximum integration time was $10^5M$ for the anti-Larmor case and $2\times 10^4M$ for the Larmor case. We chose the latter due to the existence of meta-stable orbits. The resolution of the plots in these figures is $800\times800$.

The similarity between Fig.~\ref{fig:nb} and Fig.~\ref{fig:anti-lar} (f) is striking. The main effect of the magnetic field is to distort the basin boundaries from regular lines to fractals.

Let us discuss the general structure of the basin of attraction plots and formulate the escape condition for charged particles. The main parts in the basin of attraction plots can be identified as follows:

\begin{itemize}
    \item{\bf Escape region}: The particle here escapes directly in the direction of the kick. This region is the upper right large green area in the figures. It gets reduced from left as $a$ increases. We use the boundary of this region to define an {\it effective escape energy} $\ce_{esc}$. The effective escape energy curve can be fitted with a tiny relative error to a function of the form
        \begin{equation}
        \ce_{esc}=1+\frac{a+b r_o+c r_o^2}{d+e r_o+f r_o^2}.
        \end{equation}

        where $a, b, ...\:\mbox{and}\: f$ are fitting parameters.

    \item{\bf Capture region}: This is the red nearly rectangular area in the left side of the plots when $a=0.999M$. The particle is always captured in this region for any energy. It is the proximity of $r_o$ from the horizon that makes the particle always accelerate inwardly no matter how energetic the kick is. Therefore this region shows up only when $a$ is close to $M$. Increasing $b$ for anti-Larmor motion would also lead to the emergence of this region because $r_{ms}$ would become closer to the horizon [see Fig.~\ref{rmsb} (a)].
    \item{\bf Fractal region}: The escape region is bounded by a diffuse region of fine threads that demonstrate a repetitive pattern of red, green and yellow colors. These threads get finer as they get closer to the escape region. We refer to this region as the fractal region. The particle's trajectory in it can cross the equatorial plane several times. The fractal structure is persistent at any magnification level. This fact confirms that the system is chaotic. The vertical branch of the fractal becomes smaller as $a$ decreases. The red color ceases to exist near the end of the horizontal tail of the fractal. This effect becomes more noticeable as $a$ increases until the red color completely disappears from the lower half of the fractal structure when $a=0.999M$.
    \item{\bf Meta-Stability region}: It is represented by the blue strip in the Larmor motion plots. We expect that the left boundary of this region becomes smooth if the numerical integrator is run for longer time. However, increasing the integration time will increase the computation time immensely without modifying significantly the quantity we want to measure, namely $D_f$. (See below.)
    \item{\bf Backscattering region}: It is the yellow isle located between the capture region and the upper branch of the fractal region. In the back\-scattering region the particle escapes in the direction opposite to the kick. Like the capture region, the backscattering region appears when $r_o$ is close to the horizon.
\end{itemize}

\subsection{Rotation and Chaoticness}

\begin{figure*}[ht]
\begin{center}
\ba
&&\hspace{0cm}\includegraphics[width=0.48\textwidth]{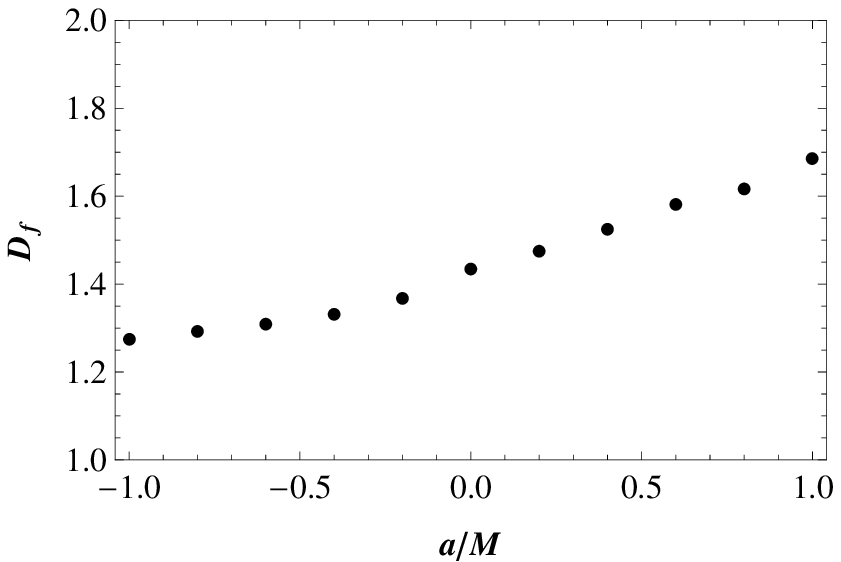}
\hspace{.2cm}\includegraphics[width=0.48\textwidth]{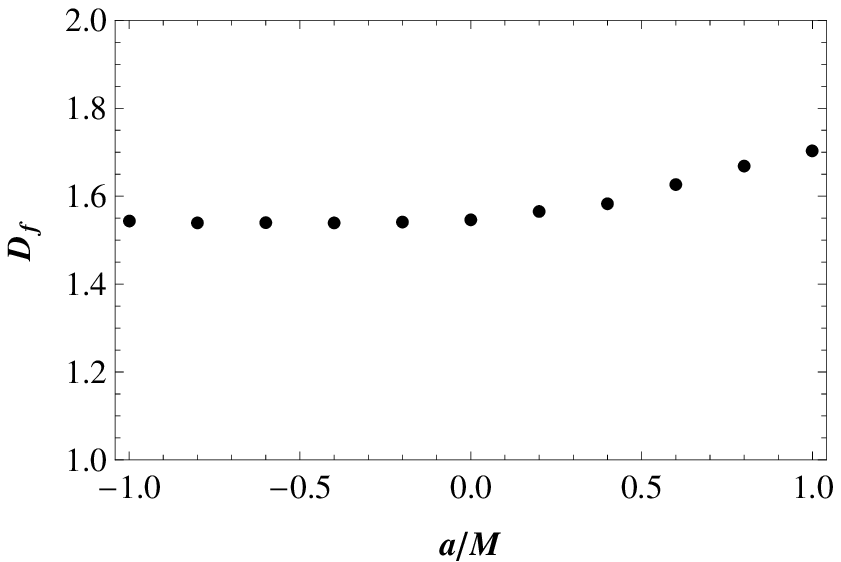}\non\\
&&\hspace{4.cm}({\bf a})\hspace{7.2cm}({\bf b})\non
\ea
\caption{The box-counting dimension $D_b$ of the basin of attraction plots of (a) Fig.~\ref{fig:anti-lar} and (b) Fig.~\ref{fig:lar} vs. $a$.}\label{fig:Df vs. a}
\end{center}
\end{figure*}

\nin It is interesting to see how the black hole's spin $a$  affects the chaoticness in the dynamics. We will use the fractal dimension $D_f$ of the basin boundary as a measure of chaoticness. The fractal dimension $D_f$ can be measured using the box-counting dimension $D_b$ which is given in a two-dimensional space of initial conditions by
\begin{equation}
D_b=\lim_{\epsilon\to 0}\left(\frac{\ln N(\epsilon)}{\ln 1/\epsilon}\right), \hspace{1cm}   1\leq D_b <2,
\end{equation}
where $N(\epsilon)$ is the number of squares of side length $\epsilon$ that are needed to cover the basin boundary~\cite{Ott}.
The box-counting dimension is related to the uncertainty exponent $\alpha\equiv 2-D_b$, which gives the probability $\rho$ that a measurement of uncertainty $\epsilon$ will fail to determine the final state of an orbit \cite{Ott}
\begin{equation}
\rho(\epsilon)\sim \epsilon^{\alpha}.
\end{equation}
It should be mentioned that $D_f$ cannot be used to make a general conclusion for the whole system since basin of attraction plots are produced for specific sets of initial conditions. Moreover, $r_{ms}$ and, to a lesser extent $\ce_o$, depend on $a$. This makes choosing consistent sets of initial conditions for different values of $a$ tricky. We preferred to take the sets of initial conditions identical to those of the basin of attraction plots in Figs.~\ref{fig:anti-lar} and~\ref{fig:lar}, in which the dependence of $r_{ms}$ on $a$ is taken into account. We kept the sets of $\ce$ unchanged for simplicity. The dependence of $D_b$ on $a$ is shown in Fig.~\ref{fig:Df vs. a} for (a) anti-Larmor motion and (b) Larmor motion.


\nin For Larmor retrograde motion $D_b$ is nearly constant while it is linearly increasing, within error, for the remaining cases. It is not surprising that $D_b$ increases with $a$ since the 'gravitational field' gets more intense as $r_{ms}$ gets closer to the horizon. As mentioned above, the $D_b$--$a$ relation depends on the set of initial conditions chosen. For example, $D_b$ becomes inversely proportional to $a$ if $r_o\in[4M,10M]$ is chosen instead.

\section{Summary}\label{sum}

We have studied the escape of neutral and charged particles kicked from circular orbits around a weakly magnetized rotating black hole. It was found that the escape of a neutral particle depends mainly on the proximity of its initial orbit to the black hole. If the particle's orbit is very close to the horizon it always gets captured. If the orbit is far enough from the horizon then it always escapes if it is made energetically free. When the orbit lies between these escape and capture regions, the particle can escape if it can be made energetically free and outwardly accelerating.

The problem is more involved for charged particles. The dynamics becomes chaotic. The final fate of a charged particle was also found to depend mainly on the initial orbit's radius. The escape and capture regions are not as lucid as in the case of a neutral particle, however. The chaoticness in the dynamics manifests itself in the boundaries between different regions of capture and escape in the space of initial conditions.

There does not seem to be an explicit general relationship between the black hole's rotation and the chaoticness in the dynamics. Instead, a restricted relationship can be given for specific sets of initial conditions. Nonetheless, the dynamics appears to be more chaotic near the black hole's horizon where the gravitational field is stronger.

It would be interesting to see how the problem develops when further sophistications are involved. Namely, when more realistic magnetic fields, more general initial orbits and more general kicks with physical kicking mechanisms are used. While these modifications may enrich the problem, we expect its main features to be sustained.

\appendix*
\begin{widetext}

\section{$r$ and $\theta$ components of the dynamical equation}

In this appendix we write the $r$ and $\theta$ components of the dynamical equation (\ref{de}). They respectfully read
\begin{eqnarray}
\ddot{r}=&&\frac{M\Delta(2r^2-\Sigma)}{\Sigma^3}\left(2a\sin^2\theta \:\dot{t}\dot{\phi}-\dot{t}^2\right)+\frac{r\Delta}{\Sigma}\dot{\theta}^2-\left(\frac{r}{\Sigma}-\frac{\Delta_{,r}}{2\Delta}\right)\dot{r}^2- \frac{\Sigma_{,\theta}}{\Sigma}\dot{r}\dot{\theta}-\frac{\Delta(2rA-\Sigma A_{,r})\sin^2\theta}{2\Sigma^3}\dot{\phi}^2 \nonumber\\
&+&2b\left[\frac{aM\Delta(2r^2-\Sigma)\sin^2\theta}{\Sigma^3}\dot{t}-\frac{\Delta(2rA-\Sigma A_{,r})\sin^2\theta}{2\Sigma^3}\dot{\phi}\right],\label{req}\\
\ddot{\theta}=&-&\frac{2r}{\Sigma}\dot{r}\dot{\theta}-\frac{Mr\Sigma_{,\theta}}{\Sigma^3}\dot{t}^2-\frac{\Sigma_{,\theta}}{2\Sigma}\dot{\theta}^2+
\frac{\Sigma_{,\theta}}{2\Delta\Sigma}\dot{r}^2-\frac{2aMr(2\Sigma\cos{\theta}-\Sigma_{,\theta}\sin{\theta})\sin{\theta}}{\Sigma^3}\dot{t}\dot{\phi}\non\\
&-&\frac{[A(\Sigma_{,\theta}\sin{\theta}-2\Sigma\cos{\theta})-\Sigma A_{,\theta}\sin{\theta}]\sin{\theta}}{2\Sigma^3}\dot{\phi}^2 \non \\
&-&2b{\bigg\{}\frac{aMr(2\Sigma\cos{\theta}-\Sigma_{,\theta}\sin{\theta})\sin{\theta}}{\Sigma^3}\dot{t}+\frac{[A(\Sigma_{,\theta}\sin{\theta}-2\Sigma\cos{\theta})-\Sigma A_{,\theta}\sin{\theta}]\sin{\theta}}{2\Sigma^3}\dot{\phi}{\bigg\}},\label{teq}
\end{eqnarray}
where $\dot{t}$ and $\dot{\phi}$ are eliminated using the expression for $\cl$ and the normalization condition $u^{\mu}u^{\nu}g_{\mu\nu}=-1$.

\end{widetext}

\end{document}